\documentclass[twocolumn,preprintnumbers,amsmath,amssymb,superscriptaddress,nofootinbib,longbibliography]{revtex4-1}

\usepackage{graphicx}
\usepackage{float}
\usepackage{bm}
\usepackage{bbm}
\usepackage{bbold}
\usepackage{dsfont}
\usepackage[usenames,dvipsnames]{xcolor}
\usepackage{pstricks}
\usepackage[tight]{subfigure}
\usepackage{verbatim}
\usepackage{units}
\usepackage{multirow}
\usepackage{enumitem}
\usepackage{mathrsfs}
\usepackage{leftidx}
\usepackage{xspace}

\usepackage[a4paper]{hyperref}
\hypersetup{colorlinks=true,linktoc=all,linkcolor=Blue,breaklinks=true,citecolor=Blue,urlcolor=Blue}

\usepackage[english]{babel}
\newcommand{\via}{\emph{via}\xspace}
\newcommand{\etal}{\emph{et al.}\xspace}

\newcommand{\kB}{k_\text{B}}

\newcommand{\intd}{\text{d}}

\newcommand{\op}[1]{\hat{#1}}

\newcommand{\dVg}{\delta V_{\text{g}}}

\newcommand{\dVgeffo}{\delta V_{\text{g}}^\text{eff}}

\newcommand{\Vg}{V_\text{g}}
\newcommand{\Kmax}{K_\text{max}}

\newcommand{\dUc}{\delta U^{\text{(c)}}}
\newcommand{\dUg}{\delta U^{\text{(g)}}}
\newcommand{\dUeta}{\delta U^{(\eta)}}
\newcommand{\dUbareta}{\delta U^{(\overline{\eta})}}

\newcommand{\dQeta}{\delta Q^{(\eta)}}

\newcommand{\LL}{\text{L}}
\newcommand{\RR}{\text{R}}
\newcommand{\GG}{\text{g}}
\newcommand{\CC}{\text{c}}

\newcommand{\elec}{\text{e}}
\newcommand{\hole}{\text{h}}

\newcommand{\perT}{\mathcal{T}}

\newcommand{\tauc}{\tau_{\text{c}}}
\newcommand{\taug}{\tau_{\text{g}}}
\newcommand{\taueta}{\tau_\eta}
\newcommand{\tauRCo}{\tau_{RC}}

\newcommand{\matSF}[1]{\underline{\bm{\mathcal{S}}}_{#1}}
\newcommand{\matSFel}[1]{\mathcal{S}_{#1}}

\newcommand{\Cc}{\mathcal{C}^{(\text{c})}}
\newcommand{\Cg}{\mathcal{C}^{(\text{g})}}
\newcommand{\cVg}{\mathcal{V}}

\newcommand{\Cqc}{C_{\text{q}}^{(\text{c})}}
\newcommand{\Cqg}{C_{\text{q}}^{(\text{g})}}
\newcommand{\Cqeta}{C_{\text{q}}^{(\eta)}}
\newcommand{\Ctot}{C_{\mu}}
\newcommand{\Ctoto}{C_{\mu}}

\newcommand{\Rtoto}{R}

\newcommand{\Ltoto}{L}

\newcommand{\chieta}{\chi^{(\eta)}}
\newcommand{\alphaeta}{\alpha^{(\eta)}}
\newcommand{\alphac}{\alpha^{(\text{c})}}
\newcommand{\alphag}{\alpha^{(\text{g})}}

\newcommand{\Fg}{\mathcal{F}}
\newcommand{\Fgeff}{\mathcal{F}^\text{eff}}

\newcommand{\Fhar}{\mathcal{F}_\text{har}}
\newcommand{\FLor}{\mathcal{F}_\text{Lor}}
\newcommand{\fLor}{F_\text{Lor}}
\newcommand{\Fsbox}{\mathcal{F}_\text{s-box}}

\newcommand{\Npart}{\mathcal{N}}

\newcommand{\tp}{t_{\text{p}}}
\newcommand{\tpk}{t_{\text{p}}^{(k)}}

\newcommand{\Iad}{I^{(1)}}

\newcommand{\vd}{v_{\text{d}}}

\newcommand{\avQad}{\overline{\mathcal{Q}}^{\,\raisebox{-2pt}{\scriptsize$(1)$}}}
\newcommand{\avQ}{\overline{\mathcal{Q}}}

\newcommand{\dNeh}{\Delta\mathcal{N}_\text{eh}}

\let\stdforall\forall
\renewcommand{\forall}{\mathop{\vphantom{\sum}\mathchoice
  {\hbox{\scalebox{1.5}{$\stdforall$}}}
  {\hbox{\scalebox{1.5}{$\stdforall$}}}{\stdforall}{\stdforall}}\displaylimits}

\usepackage[normalem]{ulem}

\newcommand{\new}[1]{\color{black}#1\color{black}\xspace} 

\begin{document}


\title{Shaping charge excitations in chiral edge states with a time-dependent gate voltage}

\author{Maciej Misiorny}
\email{misiorny@amu.edu.pl}
\affiliation{Department of Microtechnology and Nanoscience MC2, Chalmers University of Technology, SE-412 96 G\"{o}teborg, Sweden}
\affiliation{Faculty of Physics, Adam Mickiewicz University, 61-614 Pozna\'{n}, Poland}
\author{Gwendal F\`{e}ve}
\affiliation{Laboratoire Pierre Aigrain, Ecole Normale Sup\'{e}rieure-PSL Research University, CNRS, Universit\'{e} Pierre et Marie Curie-Sorbonne Universit\'{e}s, Universit\'{e} Paris Diderot-Sorbonne Paris Cit\'{e}, F-75231 Paris Cedex 05, France}
\author{Janine Splettstoesser}
\email{janines@chalmers.se}
\affiliation{Department of Microtechnology and Nanoscience MC2, Chalmers University of Technology, SE-412 96 G\"{o}teborg, Sweden}

\date{\today}

\begin{abstract}
We study a coherent conductor supporting a single edge channel in which alternating current pulses are created by local time-dependent gating and sent on a beam-splitter realized by a quantum point contact. The current response to the gate voltage in this setup is intrinsically linear. Based on a fully self-consistent treatment employing a Floquet scattering theory, we analyze the effect of different voltage shapes and frequencies, as well as the role of the gate geometry on the injected signal. In particular, we highlight the impact of frequency-dependent screening on the process of shaping the current signal. The feasibility of creating true single-particle excitations with this method is confirmed by investigating the suppression of excess noise, which is otherwise created by additional electron-hole pair excitations in the current signal.
\end{abstract}



\maketitle

\section{Introduction}

The controlled injection of single-electron excitations into electronic conductors~\cite{singleelectron_pssb} is an essential prerequisite in various research fields ranging from metrology~\cite{Pekola_Rev.Mod.Phys.85/2013} to the emerging field of quantum optics with electrons~\cite{Bocquillon_Ann.Phys.526/2014}. 
So far, two different approaches allowing for the realization of such single-electron sources have been proposed and experimentally verified.
The first approach exploits the discrete level spectrum or Coulomb blockade effects in strongly confined systems, guaranteeing that particles are emitted consecutively
when the device is subject to a time-dependent driving potential. The~feasibility of this method has been successfully demonstrated, for example, in mesoscopic capacitors realized in the quantum Hall regime~\cite{Feve_Science316/2007,Bocquillon_Phys.Rev.Lett.108/2012,Bocquillon_Science339/2013}, in superconducting turnstiles~\cite{Maisi_NewJ.Phys.11/2009}, dynamical quantum dots~\cite{Blumenthal_Nat.Phys.3/2007,Leicht_Semicond.Sci.Technol.26/2011,Giblin_Nat.Commun.3/2012}
or by using sound waves to expel electrons from a~dot~\cite{Hermelin_Nature477/2011,McNeil_Nature477/2011}.
The second \mbox{---completely different---} approach resorts to a~specific shaping of a time-dependent bias voltage applied across a junction in an otherwise unconfined conductor, resulting in the creation of so-called \emph{levitons}~\cite{Keeling_Phys.Rev.Lett.97/2006,Levitov_J.Math.Phys.37/1996,Keeling_Phys.Rev.Lett.101/2008,Dubois_Nature502/2013,Gabelli_Phys.Rev.B87/2013,Rech_Phys.Rev.Lett.118/2017}. 
However, a~matter that remains unresolved with both approaches is the \textit{local} creation of single-electron excitations in systems or materials where a strong size-confinement cannot be achieved. For instance, this is the case in topological insulators~\cite{Inhofer_Phys.Rev.B88/2013,Hofer_Phys.Rev.B88/2013,Hofer_Europhys.Lett.107/2014,Xing_Phys.Rev.B90/2014,Calzona_Phys.Rev.B94/2016,Mueller_Phys.Rev.B95/2017,Acciai_Phys.Rev.B96/2017}, in which quantum optics experiments with helical edge states have been proposed~\cite{Ferraro_Phys.Rev.B89/2014}, however, the realization of well-controlled quantum point contacts~(QPCs) still remains a challenge.
 
\begin{figure}[b]
	\includegraphics[width=0.65\columnwidth]{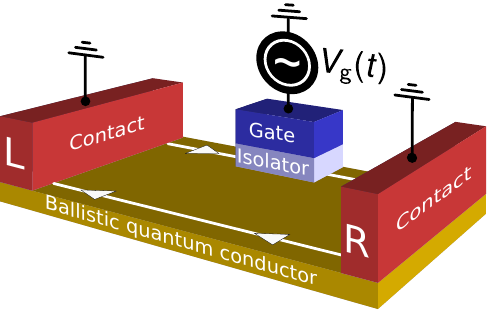}
	\caption{
	Sketch of a two-dimensional conductor with left and right (source and drain) contacts, supporting transport along two counter-propagating chiral edges. The potential landscape is locally modified by a time-dependently modulated gate voltage~$\Vg(t)$.
	}
	\label{fig1}
\end{figure}

The goal of this paper is to put forward and discuss viable schemes for a controlled production of  single-electron (and hole) pulses based on local time-dependent gating of a transport channel. 
For this purpose, we investigate a model setup consisting of a coherent conductor with a \emph{single}, chiral transport channel propagating along each edge of the sample. The conductor is locally, capacitively coupled to a gate, to which a time-dependent gate voltage is applied, see Fig.~\ref{fig1}. Such a coupling,
\new{%
arising due to Coulomb interaction between electrons in the gate and in the conductor,%
}
induces a time-dependent change of the potential landscape in the conductor. This leads, in turn, to the generation of a pure AC current response in the chiral edge channel \mbox{---described} here within the Floquet scattering matrix  approach~\cite{Pretre_Phys.Rev.B54/1996,Pedersen_Phys.Rev.B58/1998,Moskalets_Phys.Rev.B66/2002,Moskalets_Phys.Rev.B69/2004,Moskalets_book}. By treating the complex internal potentials created by the gate-voltage modulation fully self-consistently~\cite{Buttiker_Phys.Rev.Lett.70/1993,Buttiker_IlNuovoCimentoB110/1995,Buttiker_NATOASI326/1996,Buttiker_NATOASI345/1997}, we relate these internal potentials to the originally applied signals, demonstrating thereby the importance of screening for the single-particle injection scheme, especially when the driving frequency is large.

Our analysis of the time-dependent current as a function of the externally applied potential in all frequency regimes provides a clear recipe for the appropriate design of the gate-driving, which allows for the creation of 
\new{%
integer charge-current pulses well-separated in time.}\footnote{%
\new{Note that if one assumes a constant drift velocity in the vicinity of the Fermi energy, the localization in time can be actually directly mapped to a localization in space.}}  
In particular, we show that the speed at which the gate is driven is a key factor affecting the quality of the single-particle excitations created in the conductor: 
For a slow gate driving, the time-dependent current is proportional to the derivative of the applied potential (pure capacitive response), \mbox{$I(t)\propto\partial\Vg(t)/\partial t$}, which essentially means that one can shape the derivative of a gate voltage $\Vg(t)$ to obtain the desired current signal. On the contrary, in the fast-driving regime, the current has a similar shape as the potential itself, in the first half period of the driving, followed by the same signal with the opposite sign. 

Lorentzian current pulses of opposite polarity, resulting from either of the different driving schemes described above, can carry noiseless excitations of integer charges, as expected from the study of levitons~\cite{Keeling_Phys.Rev.Lett.97/2006,Levitov_J.Math.Phys.37/1996,Keeling_Phys.Rev.Lett.101/2008,Dubois_Nature502/2013,Gabelli_Phys.Rev.B87/2013}. However, the desired noiseless feature can only be tested, when the signal is partitioned at a scatterer. In the setup under consideration, due to the large separation of the edge channels propagating on opposite sides of the sample, the time-dependent modulation of the gate voltage does not induce any backscattering inherently. It rather influences the phase of the electronic wave functions in a nontrivial, time-dependent manner, resulting in the current signal described above. 
Thus, to carry out the noise analysis of the properties of emitted pulses, we assume that the time-dependent current signal induced by the gate subsequently impinges on a QPC with a finite reflection probability. In our setup, this QPC 
\new{%
with energy-independent transmission~$D$%
}
is essential for the analysis of the injected signal, since it reveals the granularity of the charge. Indeed, the charge-current noise~\cite{Mahe_Phys.Rev.B82/2010,Parmentier_Phys.Rev.B85/2012} at the barrier, is expected to be a measure of the amount of additional, spurious electron-hole pairs~\cite{Vanevic_Phys.Rev.Lett.99/2007,Vanevic_Phys.Rev.B78/2008,Vanevic_Phys.Rev.B86/2012,Vanevic_Phys.Rev.B93/2016} that are created by the time-dependent driving, limiting the accuracy with which current pulses carrying integer charge are created. Here, we show the conditions for which the excess noise vanishes both in the slow and fast driving regimes, suggesting that quantized charge-emission by local gating is feasible following the prescriptions presented in this paper.

This paper is structured as follows: In Sec.~\ref{sec:Model}, we introduce the model, the sought transport quantities of interest and the underlying theoretical approach. All results are gathered in Sec.~\ref{sec:Discussion}, where we discuss the explicit expressions for the self-consistent internal potentials~(Sec.~\ref{sec:Model_single_channel}), and the resulting current response~(Sec.~\ref{sec:Current_response}). Next, we analyze the possibility of single-particle emission both in the slow~(Sec.~\ref{sec:Adiabatic_driving}) and in the arbitrary-frequency regime~(Sec.~\ref{sec:Non-adiabatic_driving}). Finally, the characteristics of the emitted pulses are investigated from the point of view of the associated charge noise~(Sec.~\ref{sec:Noise}).

\section{\label{sec:Model}Model and formalism}

We consider a device consisting of a coherent \emph{ballistic} conductor supporting chiral edge states which locally can be subject to a time-dependent electric field generated by a gate driven by a periodic, but otherwise arbitrary potential with frequency~$\Omega$. The conductor is also attached to two metallic contacts which serve as reservoirs of electrons. 
In fact, such a setup is conceptually similar to the ``open'' mesoscopic capacitor studied by Litinski \etal~\cite{Litinski_Phys.Rev.B96/2017}, where it was shown that its current-response is intrinsically linear~\cite{Cuniberti_Phys.Rev.B57/1998}. 
In Fig.~\ref{fig1}(a) we show an example of a device in which only one edge channel is affected by the gate.
Importantly, we focus exclusively on the interaction effects due to the capacitive coupling to the gate. We neglect here interactions between electrons in the single edge channel under consideration, 
\new{%
see e.g. Ref.~\cite{Ji_Nature422/2003}.
}
In addition, we also assume that the conductor is wide enough to disregard the inter-edge electrostatic coupling, or in other words, that the coupling to the gate efficiently screens the inter-edge interactions~\cite{Gabelli_Phys.Rev.Lett.98/2007,Feve_PhysicaE76/2016}.
%
\new{%
Finally, we note that if more edge channels were involved, one would  expect the inter-edge Coulomb interaction  to renormalize the velocity of charge propagation~\cite{Sukhodub_Phys.Rev.Lett.93/2004,Degiovanni_Phys.Rev.B81/2010,Bocquillon_NatureCommun.4/2013,Grenier_Phys.Rev.B88/2013,Kamata_NatureNanotechnol.9/2014,Marguerite_Phys.Rev.B94/2016,Washio_Phys.Rev.B93/2016,Hashisaka_NaturePhys.13/2017}. This is, however, not part of a single-edge channel setup treated in this work.
}

\begin{figure}[t]
	\includegraphics[width=0.99\columnwidth]{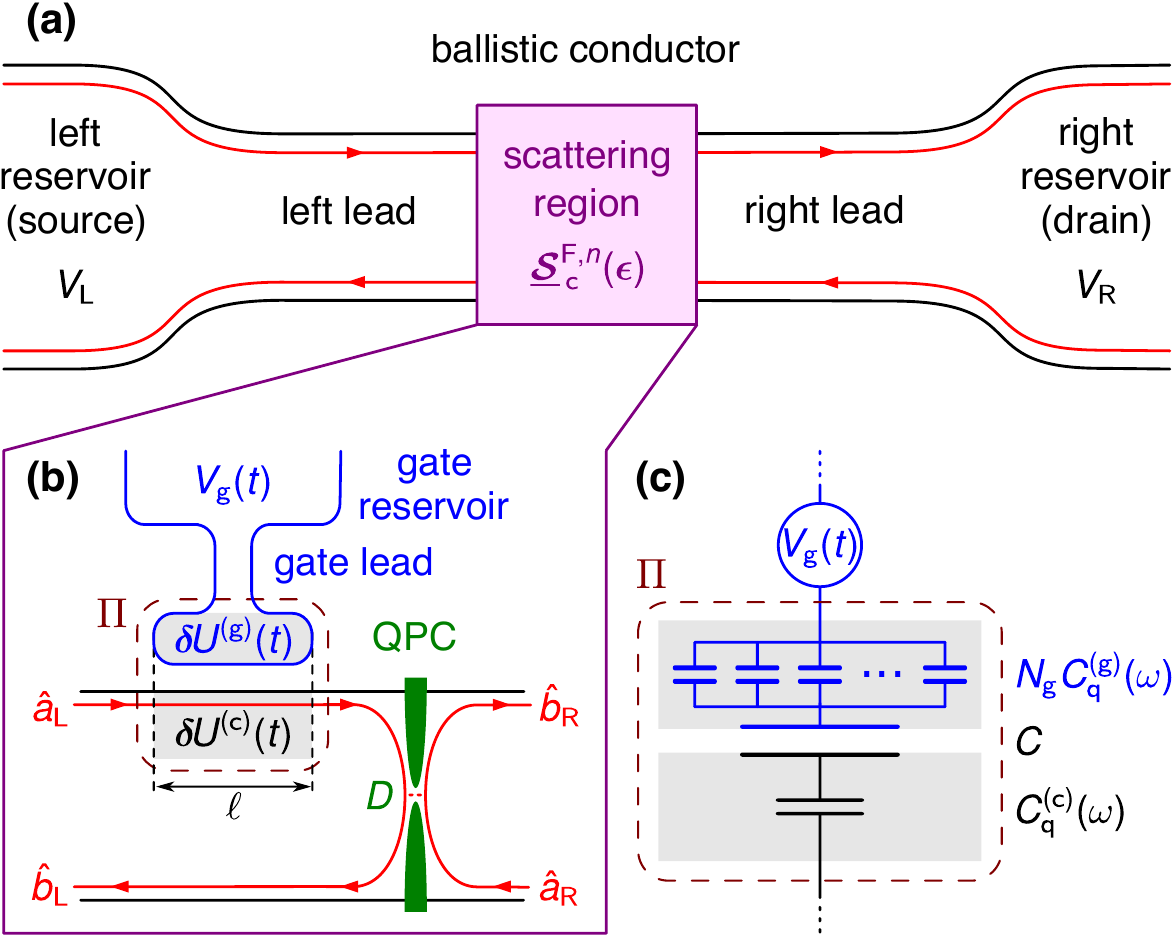}
	\caption{
	(a) Schematic illustration of a ballistic conductor supporting two counter-propagating edge channels that is connected to electronic reservoirs \via ideal leads. The effect of a time-dependently driven gate and a partitioner is generally included in the scattering region. 
	(b) An example of the scattering region for the case where the gate is coupled to only one (top) edge channel in the presence of a quantum point contact (QPC) characterized by the 
	\new{%
	energy-independent%
	}
	transmission~$D$. 
	The operators~$\op{a}_{\LL(\RR)}$ and~$\op{b}_{\LL(\RR)}$ represent annihilation of the incoming and outgoing lead states, respectively, in the left~(right) side of the conductor. 
	The long-dashed line denotes a surface~$\Pi$ enclosing a volume to which electrostatic interactions between the conductor and the gate are restrained,
	\new{
	where $\dUeta(t)$ indicates internal potentials in the gate~(\mbox{$\eta=\text{g}$}) and the conductor~(\mbox{$\eta=\text{c}$}), Eq.~(\ref{eq:dUc_A_gen}). 
	(c)~Representation of the interacting region in (b) in terms of capacitive couplings, with~$C$ denoting a purely electrostatic (geometric) capacitance and~$\Cqeta(\omega)$  standing for frequency-dependent quantum capacitances, Eq.~(\ref{eq:Cqeta_def}). Such an effective arrangement of capacitances corresponds to the total (electrochemical) capacitance~$\Ctot(\omega)$, Eq.~(\ref{eq:Ctot1}).%
	}
	For further details see Sec.~\ref{sec:Model_single_channel}.  
	}
	\label{fig2}
\end{figure}

We describe transport of electrons through this system within the Floquet scattering theory~\cite{Pretre_Phys.Rev.B54/1996,Pedersen_Phys.Rev.B58/1998,Moskalets_Phys.Rev.B66/2002,Moskalets_Phys.Rev.B69/2004,Moskalets_book}, where the electrostatic effect of the gate on electrons in the conductor is included in the model self-consistently~\cite{Buttiker_Phys.Rev.Lett.70/1993,Buttiker_IlNuovoCimentoB110/1995,Buttiker_NATOASI326/1996,Buttiker_NATOASI345/1997}.
In general, the conductor is represented here as a scattering region which is connected to reservoirs \via ideal ballistic leads, as schematically depicted in Fig.~\ref{fig2}(a). Specifically, in the situation under discussion backscattering is generally absent and it can occur only due to the presence of a partitioner, for example a quantum point contact (QPC) [see Fig.~\ref{fig2}(b)]. Furthermore, the effect of a gate, modeled as a mesoscopic capacitor~\cite{Buttiker_J.Phys.:Condens.Matter5/1993,Buttiker_Z.Phys.B94/1994},  manifests as an energy-dependent phase acquired by electrons while traversing a conductor region with the internal potential~$\dUc(t)$. The latter arises due to a capacitive coupling to the gate ---with no electron tunneling between the conductor and the gate being permitted. We note that also electrons in the gate are subject to the internal potential~$\dUg(t)$. In Fig.~\ref{fig2}(b) both these  interaction regions are  indicated as shaded areas. At this point, the  potentials~$\dUc(t)$ and~$\dUg(t)$ are taken \emph{a priori}, and they will be later derived self-consistently. Note that the conductor and the gate are treated on the same footing in the theoretical approach employed here.
Finally, the dynamics of the device is determined by the characteristic charge-relaxation time ---the $RC$-time. As shown in Sec.~\ref{sec:Adiabatic_driving}, several time-scales associated with transport of electrons contribute, in principle,  to this~$RC$-time: the~traversal time~$\tauc$ it takes an electron to pass through the interacting region of length~$\ell$ in the conductor, the time~$\taug$ an electron spends in the capacitor plate of the gate, and the time scale~\mbox{$\tau=C/g_0$} given by the purely geometric capacitance $C$.
\new{%
Here, \mbox{$g_0=e^2/h$} denotes the~quantum of conductance per spin, with $e$ standing for the electron charge (defined as negative, \mbox{$e<0$}). 
}

The incoming
\new{%
$(\op{a}_\alpha)$%
}
and outgoing
\new{%
$(\op{b}_\alpha)$%
}
states in the left and right contacts
\new{%
of the conductor in Fig.~\ref{fig2}(b)%
}
[%
\new{%
labeled%
}
by the `side' index \mbox{$\alpha=\LL(\text{eft}),\RR(\text{ight})$}], are related to each other by means of a Floquet scattering matrix. 
\new{%
In the energy representation, the annihilation operators~$\op{b}_\alpha(\epsilon)$ for outgoing states  at energy~$\epsilon$ can be related to the annihilation operators~$\op{a}_\alpha(\epsilon_n)$ for incoming states at energy~$\epsilon_n\equiv\epsilon+n\hbar\Omega$ \via the Floquet scattering matrix~$\new{\matSF{n}^{(\CC)}}(\epsilon)$ \cite{Moskalets_Phys.Rev.B66/2002,Moskalets_Phys.Rev.B69/2004,Moskalets_book} as follows
\begin{equation}\label{eq:bSa_def_c}
\renewcommand{\arraystretch}{1.35}
	\begin{pmatrix}
	\op{b}_\LL(\epsilon)
	\\
	\op{b}_\RR(\epsilon)
	\end{pmatrix}
	=
	\sum\limits_{n=-\infty}^{+\infty}
	\matSF{n}^{(\CC)}(\epsilon)
	\begin{pmatrix}
	\op{a}_\LL(\epsilon_n)
	\\
	\op{a}_\RR(\epsilon_n)
	\end{pmatrix}
	.
\end{equation}
Here,
}
$n$ quantifies the number of so-called Floquet energy quanta  $\hbar\Omega$ that an electron can emit (\mbox{$n>0$}) or absorb (\mbox{$n<0$}) during the scattering event. Recall that we consider here a case, where no backscattering takes place in the gate-driven region, which is justified by the large distance between the edge states. Therefore, only the \textit{frozen} scattering matrix enters Eq.~(\ref{eq:bSa_def_c}), in contrast to the case with backscattering in the time-dependent potential~\cite{Burmeister_Phys.Rev.B57/1998,Li_Phys.Rev.B60/1999,Hara_Phys.Rev.A62/2000} \mbox{---for} comprehensive derivations of this (frozen) scattering matrix see Refs.~\cite{Moskalets_book,Inhofer_Phys.Rev.B88/2013}. Furthermore, note that in the present discussion we assume that an edge channel corresponds to only one spin channel, and consequently, to a single transport channel (\mbox{$N_\LL=N_\RR=1$}), depicted in Fig.~\ref{fig2} as an arrowed line.
An analogous relation to Eq.~(\ref{eq:bSa_def_c}) holds for the gate except that now we allow $N_\GG$ transport channels (both orbital and spin ---not indicated in Fig.~\ref{fig2}) to participate,
\begin{equation}\label{eq:bSa_def_g}
	\op{b}_{\GG,j}(\epsilon)
	=
	\sum\limits_{j^\prime=1}^{N_\GG}
	\sum\limits_{n=-\infty}^{+\infty}
\new{
	\matSFel{n,jj^\prime}^{(\GG)}(\epsilon)
	\,
	\op{a}_{\GG,j^\prime}(\epsilon_n)
	,
}
\end{equation}
\new{%
where the operators and the scattering matrix are now additionally labeled with channel indices~$j$ and~$j^\prime$.}
This generalization allows for the treatment of metallic gates as well as of gates consisting of a similar structure with chiral edges as the conductor itself.

The general expression for the
\new{%
operator describing the%
}
total current in the 
\new{%
any lead~$\alpha=\LL,\RR,\GG$ as a function of time%
}
takes the form~\cite{Buttiker_Phys.Rev.B46/1992}:
\begin{multline}\label{eq:opI_def}
    \op{I}_\alpha(t)
    =
    \frac{e}{h}
    \sum\limits_{j=1}^{N_\alpha}
    \,
    \int\limits_{-\infty}^{+\infty}
    \!\!
    \intd\epsilon\intd\epsilon^\prime
    \,
    \text{e}^{i(\epsilon-\epsilon^\prime)t/\hbar}
\\[-0pt]
	\times   
    \Big[
    \op{b}_{\alpha,j}^\dagger(\epsilon)
    \op{b}_{\alpha,j}^{}(\epsilon^\prime)
    -
    \op{a}_{\alpha,j}^\dagger(\epsilon)
    \op{a}_{\alpha,j}^{}(\epsilon^\prime)
    \Big]
    .
\end{multline}
Importantly, 
\new{%
this operator is%
}
associated  with the flow of electrons into the direction of reservoirs.
Employing the above current operator, in the next sections we will be able to derive the experimentally relevant quantities characterizing the electronic transport properties of the system, such as:
\begin{itemize}[leftmargin=*]
\item 
The 
\new{%
expectation value~$I_\alpha(t)$ of the%
}
charge current 
\new{%
as a function of time,%
}
%
\begin{equation}\label{eq:It_def}
	I_\alpha(t)
	=
	\langle\op{I}_\alpha(t)\rangle
	,
\end{equation}
with $\langle\ldots\rangle$ denoting the quantum statistical average, and its Fourier transform defined as
\begin{equation}\label{eq:Iomega_def}
	I_\alpha(\omega)
	=
	\int\limits_{-\infty}^{\infty}
	\!\!
	\intd\omega
	\,
	\text{e}^{i\omega t}
	\,
	I_\alpha(t)
	.
\end{equation}
\item
The 
\new{%
zero-frequency%
}
charge-current noise $\mathcal{P}_{\alpha\beta}$, in the following shortly referred to as the `current noise', will be used to characterize the precision of the injected current signal~\cite{Martin_Phys.Rev.B45/1992,Blanter_Phys.Rep.336/2000},
\new{%
\begin{equation}\label{eq:Noise_def}
	\mathcal{P}_{\alpha\beta}
	=
	\frac{1}{2}
	\int
	\!\!
	\intd t^\prime
	\!
	\int\limits_0^\perT
	\!
	\frac{\intd t}{\perT}
	\,
	\langle
	\{
	\Delta\op{I}_\alpha(t+t^\prime)
	,
	\Delta\op{I}_\beta(t)
	\}
	\rangle
	.
\end{equation}
}\hspace*{-16pt}
In essence, it corresponds to the 
\new{%
zero-frequency%
}
Fourier transform of the current-current correlator time-averaged in the absolute time~$t$ over one driving cycle~\cite{Parmentier_Phys.Rev.B85/2012}, 
\new{%
with~\mbox{$\perT=2\pi/\Omega$} denoting the period of the driving,}
and $\Delta\op{I}_\alpha(t)=\op{I}_\alpha(t)-\langle\op{I}_\alpha(t)\rangle$.
\end{itemize}

To complete calculations of the current~(\ref{eq:It_def}) and the noise~(\ref{eq:Noise_def}), one eventually needs to evaluate the quantum statistical averages of 
\new{%
pairs of%
}
the operators for incoming states in leads.
For the conductor, at the ends of which only constant (i.e., time-independent) potentials~$V_\alpha$ are externally applied, these averages are basically determined by the equilibrium statistical distribution of electrons in the reservoirs~\cite{Buttiker_Phys.Rev.B46/1992},
\begin{equation}\label{eq:StatAv_c}
	\langle
	\op{a}^\dagger_{\alpha}(\epsilon)
	\op{a}^{}_{\beta}(\epsilon^\prime)
	\rangle	
	=
	\delta_{\alpha\beta}
	\delta(\epsilon-\epsilon^\prime)
	f_\alpha(\epsilon)
	\quad
	\text{for}
	\quad
	\alpha,\beta=\LL,\RR
	,
\end{equation}
where
$
	f_\alpha(\epsilon)
	=
	\{1+\exp[(\epsilon-eV_\alpha)/(\kB T)]\}^{-1}
$
is the Fermi-Dirac function, with~$T$ denoting the electronic temperature of the reservoirs and~$\kB$ being the Boltzmann constant. Note that the energy reference scale is assumed such that the equilibrium electrochemical potential of an unbiased (grounded) conductor corresponds to zero.

When a time-dependent driving bias is applied to the gate, the task becomes more complex. The driving affects the electronic wave functions in the gate reservoir, inducing a spread in energy~\cite{Battista_Phys.Rev.B90/2014}. One has to relate these reservoir states to the lead states, which in turn enter the current operator~(\ref{eq:opI_def}). This relation is given by~\cite{Pretre_Phys.Rev.B54/1996,Pedersen_Phys.Rev.B58/1998,Battista_Phys.Rev.B90/2014}
\begin{multline}\label{eq:StatAv_g}
	\langle
\new{
	\op{a}^\dagger_{\GG,j}(\epsilon)
	\op{a}^{}_{\GG,j^\prime}(\epsilon^\prime)
}
	\rangle	
	=
\new{
	\delta_{jj^\prime}
}
	\!
	\sum\limits_{n,m=-\infty}^{+\infty}
	\!
	\cVg_n^\ast
	\,
	\cVg_{n+m}^{}
\\
	\times
	\delta(\epsilon_m-\epsilon^\prime)
	f_\GG(\epsilon_{-n})
	,
\end{multline}
with
\begin{equation}\label{eq:cVg}
\hspace*{-5pt}
	\cVg_n
	=
	\int\limits_0^\perT
	\!
	\frac{\intd t}{\perT}
	\, 
	\text{e}^{in\Omega t}
	\text{e}^{-i\varphi_\text{g}(t)}
	\ \,
	\text{and}
	\ \,
	\varphi_\text{g}(t)
	=
	\frac{e}{\hbar}
	\int\limits_0^t
	\!\!
	\intd t^\prime
	\,
	\dVg(t^\prime)	
	.
	\!
\end{equation}
In the equation above, the dynamic phase~$\varphi_\text{g}(t)$ is determined entirely by the pure AC component $\dVg(t)$ of the external 
\new{%
gate potential%
}
\mbox{$V_\GG(t)=V_\GG+\dVg(t)$}, while the Fermi-Dirac distribution function
\new{of the gate}
in Eq.~(\ref{eq:StatAv_g}) depends only on the DC component~$V_\GG$.

Finally, we emphasize that~$\dVg(t)$ stands for a pure AC signal of an otherwise completely arbitrary form. We are going to make use of this in the analysis of different driving potentials in Sec.~\ref{sec:Discussion}. In experiments voltage signals of a more complex form, such as a series of Lorentizans or steps, are often constructed as a superposition of several harmonics~\cite{Dubois_Nature502/2013,Gabelli_Phys.Rev.B87/2013}.
For this purpose, let us first write the AC gate voltage as
$
	\dVg(t)
	=
	\dVg
	\Fg(t)
	,
$
with~$\Fg(t)$ representing a dimensionless, periodic and real function and~$\dVg$ denoting the magnitude of the driving voltage. One can then  express the driving function~$\Fg(t)$ as a harmonic series
\begin{equation}\label{eq:Fg_expand}
	\Fg(t)
	=
	\sum\limits_{k=1}^{\Kmax}
	\text{Re}
	\Big\{
	2\Fg_k
	\text{e}^{-ik\Omega t}
	\Big\}	
	,
\end{equation}
where~$\Kmax$ denotes the number of consecutive 
harmonics~\new{$k$} 
taken into account, and the expansion (Fourier) coefficients~$\Fg_k$ are given by 
\begin{equation}\label{eq:Fg_k_coeff}
	\Fg_k
	=
	\int\limits_0^\perT\!
	\frac{\intd t}{\perT}\,
	\text{e}^{in\Omega t}\,
	\Fg(t)
	.
\end{equation}
Note that the term~$\Fg_0$ is absent in Eq.~(\ref{eq:Fg_expand}) because it essentially corresponds to the time averaging of~$\Fg(t)$ over one period~$\perT$, which for the AC component, by definition, is equal to zero. 
Importantly, Eq.~(\ref{eq:Fg_expand}) can also be easily Fourier-trans\-form\-ed, 
\begin{equation}\label{eq:Fg_expand_FT}
	\Fg(\omega)
	=
	2\pi
	\!
	\sum_{k=-\Kmax}^{\Kmax}
	\!
	\Fg_k
	\,
	\delta(\omega-k\Omega)
	,
\end{equation}
in terms of a continuous frequency $\omega$. This expression will prove particularly convenient when it comes to the numerical evaluation of time-dependent currents due to specific driving potentials, Eqs.~(\ref{eq:I1_t_gen})-(\ref{eq:dVgeff1_t}). 

\section{\label{sec:Discussion}Results and discussion}

\subsection{\label{sec:Model_single_channel}Gate applied to a single edge channel}

In Fig.~\ref{fig2}(b), we show a sketch of the conductor, where only one (top) edge channel is capacitively coupled to the time-dependently driven gate. This is the generic situation treated in this paper.
Moreover, a quantum point contact with 
\new{
energy-independent}\footnote{%
\new{The validity of this condition can be technically checked in experiments by measuring the partition noise as a function of a DC bias, and showing that this dependence is linear \mbox{---see,} e.g., the supplementary information of Ref.~\cite{Bocquillon_Phys.Rev.Lett.108/2012}. 
}
}
transmission $D$ serves as a partitioner for the stream of impinging electrons traveling along the edges. Its presence is essential for the analysis of the nature of created charge pulses, as will be seen in Sec.~\ref{sec:Noise}.

The Floquet scattering matrix $\new{\matSF{n}^{(\CC)}}(\epsilon)$ for the conductor is given by
\begin{equation}\label{eq:SFnc_A_def}
	\renewcommand{\arraystretch}{1.75}
	\new{\matSF{n}^{(\CC)}}(\epsilon)
	=
	\begin{pmatrix}
	\sqrt{1-D}\,\Cc_{-n}\,\text{e}^{i\epsilon_n\tauc/\hbar} & \sqrt{D}\,\delta_{n0}
	\\
	\sqrt{D}\,\Cc_{-n}\,\text{e}^{i\epsilon_n\tauc/\hbar} & -\sqrt{1-D}\,\delta_{n0}
	\end{pmatrix}
	,
\end{equation}
while the elements of the gate scattering matrix~$\new{\matSFel{n,jj^\prime}^{(\GG)}}(\epsilon)$ read
\begin{equation}\label{eq:SFng_A_def}
	\new{\matSFel{n,,jj^\prime}^{(\GG)}}(\epsilon)
	=
	\new{\delta_{jj^\prime}}
	\,
	\Cg_{-n}
	\,
	\text{e}^{i\epsilon_n\taug/\hbar}
	.
\end{equation}
Here, the assumption has been made that no inter-channel electron scattering process are allowed in the gate and all transport channels are described by the same dwell time~$\taug$. These scattering matrices essentially capture the fact that when an electron passes through the region with an internal time-dependent potential being present, it acquires an additional energy-dependent phase weighted by the probability amplitude~$\mathcal{C}_{n}^{(\eta)}$ that 
\new{
$n$~Floquet energy quanta~$\hbar\Omega$ are absorbed~(\mbox{$n<0$}) or emitted~(\mbox{$n>0$}) by this electron.%
}
Specifically, for~\mbox{$\eta=\text{c(onductor)},\text{g(gate)}$}, these amplitudes take the following form 
\begin{equation}\label{eq:Ceta}
	\mathcal{C}^{(\eta)}_n
	=
	\int\limits_0^{\perT}
	\!
	\frac{\intd t}{\perT}
	\,
	\text{e}^{in\Omega t}
	\text{e}^{-i\phi^{(\eta)}(t)}
	,
\end{equation}
with the phase $\phi^{(\eta)}(t)$ defined as
\begin{equation}\label{eq:phieta}
	\phi^{(\eta)}(t)
	=
	\frac{e}{\hbar}
	\int\limits_{t-\tau_\eta}^t
	\!\!\!
	\intd t^\prime
	\,
	\dUeta(t^\prime)
	.
\end{equation}

Hitherto, we have presumed in our discussion that the internal potentials~$\dUc(t)$ and~$\dUg(t)$ are known. Since such internal potentials generally emerge in conductors as a consequence of modification of the charge distribution due to Coulomb interaction with some other neighboring \mbox{---purely} capacitively \mbox{coupled---} metallic conductors~\cite{Buttiker_Phys.Rev.Lett.70/1993,Buttiker_IlNuovoCimentoB110/1995,Buttiker_NATOASI326/1996,Buttiker_NATOASI345/1997}, one can determine them self-consistently by demanding over-all charge and current conservation. For the present system, this approach  yields (for details and further discussion see Appendix~\ref{app:Self-consist_interact})
\begin{equation}\label{eq:dUc_A_gen}
	\dUeta(t)
	=	
	\int\limits_{-\infty}^{+\infty}
	\!\!
	\intd t^\prime
	\,
	\alphaeta_1(t-t^\prime)
	\,
	\dVg(t^\prime)
	.
\end{equation}
Using explicitly the expansion of the driving potential~$\dVg(t)$ into harmonics, Eqs.~(\ref{eq:Fg_expand})-(\ref{eq:Fg_expand_FT}), this can be written as
\begin{equation}\label{eq:dUc_A}
	\dUeta(t)
	=
	\dVg
	\sum\limits_{k=1}^{\Kmax}
	\text{Re}
	\Big\{
	2\Fg_k
	\,
	\text{e}^{-ik\Omega t}
	\,
	\alphaeta(k\Omega)
	\Big\}
	,
\end{equation}
with the coefficient~$\Fg_k$ defined in Eq.~(\ref{eq:Fg_expand_FT}), and the lever arm functions
\begin{subequations}\label{eq:alpha1}
\begin{gather}\label{eq:alphac1}
	\alphac(\omega)
	=
	\frac{\Ctoto(\omega)}{\Cqc(\omega)}
	,
\\[5pt]\label{eq:alphag1}
	\alphag(\omega)
	=
	\bigg[
	\frac{1}{\raisebox{-0.5ex}{$C$}}+\frac{1}{\Cqc(\omega)}
	\bigg]
	\Ctoto(\omega)
	.
\end{gather}
\end{subequations}
The asymmetric form of these lever arm functions for the conductor~(\ref{eq:alphac1}) and the gate~(\ref{eq:alphag1}) is due the fact that we assume the driving potential to be applied exclusively to the gate. Note, however, that all results for the measurable current are gauge invariant. Furthermore,~$C_\text{q}^{(\eta)}(\omega)$ is the frequency-dependent quantum capacitance, defined~as
\begin{equation}\label{eq:Cqeta_def}
	C_\text{q}^{(\eta)}(\omega)
	=
	i
	C_\text{q}^{(\eta)}
	\frac{1-\exp(i\omega\taueta)}{\omega\taueta}
	,
\end{equation}
with~\mbox{$\Cqeta=g_0\taueta$}, and $\Ctoto(\omega)$ stands for the total (electrochemical) capacitance of a purely electrostatic capacitance~$C$ connected in series with quantum capacitances~$\Cqc(\omega)$ and~$N_\GG\Cqg(\omega)$, that is,
\begin{equation}\label{eq:Ctot1}
	\frac{1}{\raisebox{-0.5ex}{$\Ctoto(\omega)$}}
	=
	\frac{1}{\raisebox{-0.5ex}{$C$}}
	+
	\frac{1}{\Cqc(\omega)}
	+
	\frac{1}{N_\GG\Cqg(\omega)}
	.
\end{equation}
\new{See also Fig.~\ref{fig2}(c) for a schematic depiction of the equation above.%
}
Worthy of note here is that Eq.~(\ref{eq:Ctot1}) strongly depends on the choice of the gate implementation. In particular, if the gate has the form of a metallic conductor, the available number of transport channels~$N_\GG$ is very large~(\mbox{$N_\GG\rightarrow\infty$}). Consequently, one can neglect  the last term of Eq.~(\ref{eq:Ctot1}),
\new{
so that $\Cqg(\omega)$ does not enter the physics of~$\Ctot(\omega)$.%
}
Then, it follows from Eq.~(\ref{eq:alphag1}) that in such a limit $\dUg(t) =\dVg(t)$.
This differs from the case where the gate has properties similar to the conductor itself and it only supports few edge channels, or in particular, just a single channel as considered in this paper. In such a case, the last term of Eq.~(\ref{eq:Ctot1}) contributes significantly.

The self-consistent potentials in Eq.~(\ref{eq:dUc_A}), together with the following explicit equations~(\ref{eq:alpha1}) for their ingredients, provide the basis for the detailed investigation of the physical role played by the gate driving potential for the creation of charge current pulses. To conclude the present discussion, we emphasize that the approach used above (see also Appendix~\ref{app:Self-consist_interact}) is in general easily applicable only in the situation when a system exhibits a linear AC response to potentials, both external and internal. 
Importantly, here we have exploited the fact that the current response of chiral edge channels is inherently linear
\new{in the absence of backscattering,}
irrespective of the size of the applied potentials, see Eqs.~(\ref{eq:IL_t_A})-(\ref{eq:Ig_t_A}). In fact, this property is generic to one-dimensional electron systems, as first pointed out by Cuniberti~\etal~\cite{Cuniberti_Phys.Rev.B57/1998}.

\subsection{Current response}\label{sec:Current_response}

To begin the analysis of the properties of the charge pulses formed in the conductor by means of the time modulation \via the gate electrode,  we start with the discussion of the frequency-dependent charge current response~$I_{\RR}(\omega)\equiv I(\omega)$, Eq.~(\ref{eq:Iomega_def}), in the right reservoir. 
Since we are interested in the charge current-pulses created exclusively by the time-dependent gate driving, we henceforth assume that no constant voltage bias is applied across the conductor ($V_\LL=V_\RR$).  We find
\begin{equation}\label{eq:I1_omega}
	I(\omega)
	=
	G(\omega)
	\dVg(\omega)
	,
\end{equation}
with the admittance~$G(\omega)$ of the form
\begin{equation}\label{eq:G1_omega}
	G(\omega)
	=
	-
	i\omega
	D
	\Ctoto(\omega)	
	\equiv
	-
	i\omega
	DCg(\omega)
	,
\end{equation}
and the dimensionless auxiliary relaxation function~$g(\omega)$ defined~as
\begin{equation}\label{eq:g1_aux}
	\frac{1}{g(\omega)}
	=
	1
	-
	\frac{i\omega\tau}{1-\exp(i\omega\tauc)}
	-
	\frac{1}{N_\GG}
	\cdot
	\frac{i\omega\tau}{1-\exp(i\omega\taug)}
	,
\end{equation}
where $\tau=C/g_0$ is the time scale associated with a capacitive coupling between the conductor and the gate.
Note that in the limit of a metallic gate electrode~(\mbox{$N_\GG\rightarrow\infty$}) the last term of Eq.~(\ref{eq:g1_aux}) can be neglected. A a result,  in this limit, we obtain the relaxation function~$g(\omega)$ found by Mora and Le~Hur~\cite{Mora_Nat.Phys.6/2010} for the quantum analogue of an~$RC$ circuit based on a cavity connected \via a quantum point contact (QPC) to a reservoir of electrons.

First of all, from Eq.~(\ref{eq:I1_omega}) one concludes that the system under discussion exhibits inherently \emph{linear} current response to the voltage~$\dVg(\omega)$ applied at the gate. 
As mentioned above, this is a general property of a one-dimensional electron system with interactions~\cite{Cuniberti_Phys.Rev.B57/1998}.
Importantly, we emphasize that while deriving this result no constraints regarding the size of~$\dVg(\omega)$ and the driving frequency~$\Omega$ were imposed. We also note that analogous result have been recently found by Litinski~\etal~\cite{Litinski_Phys.Rev.B96/2017} for a mesoscopic capacitor
\new{%
with transmission~\mbox{$D\approx1$}}
by means of a bosonisation formalism.
\new{%
Secondly,%
}
it can be seen that the non-trivial dynamics of the current response is determined by the interplay of three time scales, cf.~Eqs.~(\ref{eq:G1_omega})-(\ref{eq:g1_aux}), set by: the traversal time~$\tauc$ for the conductor, the dwell time~$\taug$ of the gate plate and the time scale~$\tau$ related to the geometric capacitance~$C$.

The expression for the time-resolved current, which is of main interest for the present work, is obtained by performing the inverse Fourier transform of Eq.~(\ref{eq:I1_omega}), 
\begin{equation}\label{eq:I1_t_gen}
	I(t)
	=
	D\Ctoto
	\frac{\intd \dVgeffo(t)}{\intd t}
	,
\end{equation}
with~\mbox{$\Ctoto\equiv\Ctoto(\omega=0)$} and the effective gate potential~$\dVgeffo(t)$ defined as
\begin{equation}
	\dVgeffo(t)
	=
	\frac{C}{\Ctoto}
	\!
	\int\limits_{-\infty}^{+\infty}
	\!\!
	\intd t^\prime
	\,
	g(t-t^\prime)
	\dVg(t^\prime)
	.
\end{equation}
In general, an analytical derivation of the inverse transform~$g(t)$ of the~re\-laxation function~$g(\omega)$, Eq.~(\ref{eq:g1_aux}), poses a non-trivial task. For a metallic gate electrode~(\mbox{$N_\GG\rightarrow\infty$}) $g(t)$ can be conveniently obtained using the Lambert function~\cite{Litinski_Phys.Rev.B96/2017}.
On the other hand, for a generic gate, the expression for the effective gate potential can be brought to a convenient form with the help of the expansion into harmonics~(\ref{eq:Fg_expand_FT}), 
\begin{equation}\label{eq:dVgeff1_t}
	\dVgeffo(t)
	=
	\dVg
	\sum\limits_{k=1}^{\Kmax}
	\text{Re}
	\bigg\{
	2
	\Fg_k
	\,
	\text{e}^{-ik\Omega t}
	\,
	\frac{\Ctoto(k\Omega)}{\Ctoto}
	\bigg\}
	.
\end{equation}
One can, thus, immediately see that the current response of the desired shape can be achieved by proper engineering of harmonic components of the gate driving signal.
In this paper, we are going to concentrate on three different types of driving functions~$\Fg(t)$:
(i)~a~harmonic function which, representing the simplest AC driving, provides a suitable starting point for our considerations;
(ii)~a~periodic Lorentzian which, as we will see in Sec.~\ref{sec:Non-adiabatic_driving}, allows for generation of clean electron/hole pulses in the non-adiabatic limit;
(iii)~a~smooth-box function which is the natural choice to emit clean single electron/hole pulses in the slow driving limit, as discussed in Sec.~\ref{sec:Adiabatic_driving}. 
Specifically, these three driving functions are defined as follows:
\begin{enumerate}[align=parleft,leftmargin=*,labelsep=-2pt,label=(\roman*)]
\item
a \emph{harmonic} gate driving ---the dashed-dotted line in Fig.~\ref{fig3}(a),
\begin{equation}\label{eq:F_har}
	\Fhar(t)
	=
	\frac{1}{2}
	\cos\!\big(\Omega [t-\tp]\big)
	,
\end{equation}
\noindent with~$0\leqslant\tp<\perT$ denoting the position of the signal maximum within one period;
\item
a \emph{periodic Lorentzian} gate driving ---the dashed line in Fig.~\ref{fig3}(a), 
\begin{multline}\label{eq:F_L}
	\FLor(t)
	=
	-
	\frac{1}{2}
	\Big[
	\text{Im}
	\Big\{
	\!
	\cot
	\!
	\big(\Omega[t-\tp+i\Gamma]/2\big)
	\Big\}	
	+
	1
	\Big]
\\
	\times
	\sinh
	\!
	\big(\Omega\Gamma\big)
	,
\end{multline}
where~$2\Gamma$ describes the full width at half maximum (see also Appendix~\ref{app:Aux_fun} for a further analysis of this driving function);
\item 
a \emph{smooth-box} gate driving designed to yield the current~$I(t)$ in the shape of evenly distributed Lorentzians ---the solid line in Fig.~\ref{fig3}(a), 
\begin{equation}\label{eq:F_s-box}
	\Fsbox(t)
	=
	\frac{
	2F(t)
	-
	1
	}{
	2\big[
	2F(\tp)
	-
	1
	\big]
	}
	,
\end{equation}
with
\begin{equation}
	F(t)
	=
	\frac{1}{\pi}
	\text{Re}
	\bigg\{
	i
	\,
	\text{ln}
	\bigg[
	\frac{
	\sin\!\big(\Omega[t-\tp+\perT/4+i\Gamma]/2\big)
	}{
	\sin\!\big(\Omega[t-\tp-\perT/4+i\Gamma]/2\big)
	}
	\bigg]
	\!
	\bigg\}
	.
\end{equation}
\new{Similarly as for (i) and~(ii),%
}
the maximum of~$\Fsbox(t)$ within one period is located at~$\tp$, and the smearing of the steps~$\Gamma$ is chosen so that it corresponds to the half width at half maximum of Lorentzians obtained by differentiating $\Fsbox(t)$ with respect to time~$t$. 
In the limit of vanishingly small smearing~$\Gamma$, the smooth-box driving approaches the square-box potential, that is,
$
	\lim_{\Gamma\rightarrow0}
	\Fsbox(t)
	=
	\sum_{n=-\infty}^{+\infty}
	\big\{
	\theta\big[t-\tp+(n+1/4)\perT\big]
	-
	\theta\big[t-\tp+(n-1/4)\perT\big]
	\big\}
	-
	1/2
$,
which in Fig.~\ref{fig3}(a) is depicted with the finely dashed line.
\end{enumerate}
To facilitate comparison between different shapes of the gate potential, the above driving functions~(\ref{eq:F_L})-(\ref{eq:F_s-box}) have been formulated in such a way that they are characterized by the same peak-to-peak amplitude equal to~1. As one will see below, under such a condition these three types of driving yield the same average charge over a half-period in the \emph{adiabatic-response} regime \mbox{---in}~the~following referred to also as the \emph{slow driving} regime.

As will be discussed below, apart from the shape of the driving potential also the frequency at which the potential is driven plays a dominant role for the controlled emission of charge pulses. To explore this aspect, we first analyze the case of slow gate driving~(Sec.~\ref{sec:Adiabatic_driving}), which will serve later as a starting point for considerations of arbitrary-frequency driving~(Sec.~\ref{sec:Non-adiabatic_driving}).

\subsection{\label{sec:Adiabatic_driving}Slow driving (adiabatic-response) regime}

In the low-frequency regime, one can obtain analogous equations to that for a classical $RC$ circuit in series by expanding the electrochemical capacitance~$\Ctoto(\omega)$, Eq.~(\ref{eq:Ctot1}), up to the first order in frequency~$\omega$, so that the formula~(\ref{eq:G1_omega}) for the admittance~$G(\omega)$ becomes 
\begin{equation}\label{eq:G1_omega_lowfreq}
	G(\omega)
	=
	-
	i\omega
	D\Ctoto	
	\Big\{
	1
	+
	i
	\omega
	\Rtoto
	\Ctoto
	\Big\}
	.
\end{equation}
Here,
\mbox{
$
	\Rtoto
	\!
	=
	\!
	\big[
	g_0^{-1}
	\!
	+
	\!
	(N_\GG g_0)^{-1}
	\big]/2
$
} 
is the equivalent charge relaxation resistance. 
Noteworthily, for a metallic gate, that is, if~\mbox{$N_\GG\rightarrow\infty$}, this resistance reduces to~\mbox{$\Rtoto=h/(2e^2)$} \mbox{---the} well-known \emph{B\"{u}ttiker resistance}~\cite{Buttiker_Phys.Rev.Lett.70/1993,Buttiker_Phys.Lett.A180/1993,Gabelli_Science313/2006,Buttiker_In_memory}, independently of the 
temperature.\footnote{%
\new{
Note that accounting also for the next order in the frequency expansion~\cite{Wang_Phys.Rev.B75/2007,Yin_Phys.Rev.B90/2014} 
\mbox{$
	G^{(3)}
	=
	-
	i\omega^3
	D\Ctoto^2
	\big(
	\Ltoto
	-
	\Ctoto
	\Rtoto^2
	\big)
$}
brings in the equivalent quantum inductance, 
\mbox{$
	\Ltoto
	=
	\big[
	\tauc
	g_0^{-1}
	+
	\taug
	(N_\GG g_0)^{-1}
	\big]/12.
$}%
}
}

\begin{figure}[t]
	\includegraphics[scale=1]{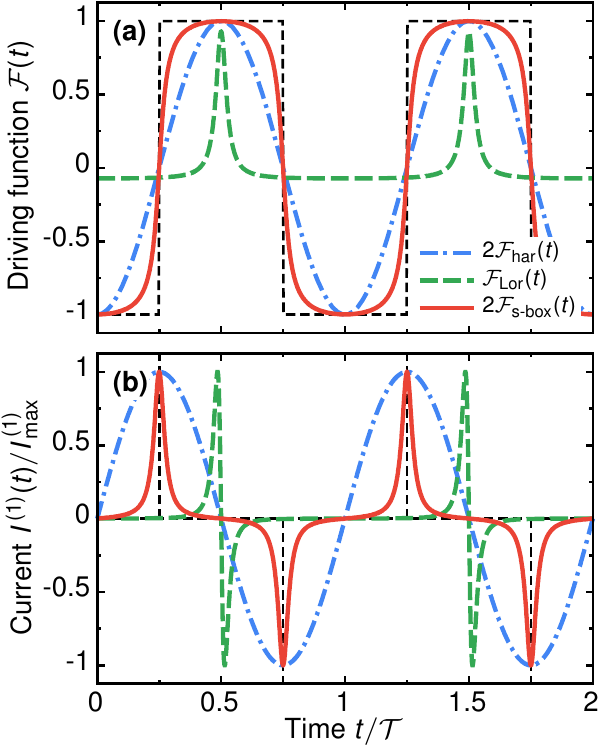}
	\caption{
	(a) Pure time-dependent (AC) component of the gate voltage $\dVg(t)=\dVg\Fg(t)$ shown for three different driving functions $\Fg(t)$: harmonic (dashed-dotted line), periodic Lorentzian (dashed line) and smooth-box (solid line), for $\tp/\perT=0.5$ and $\Gamma/\perT=0.025$. The finely dashed rectangles represent the smooth-box driving in the limit of a vanishingly small smearing~$\Gamma$.
	(b) Currents $I^{(1)}(t)$ in the adiabatic-response limit, Eq.~(\ref{eq:dVgeff1_t_ad}), corresponding to the driving potentials plotted in~(a), with~$I_{\text{max}}^{(1)}$ being the amplitude of the current. Note that
	\new{%
	the factor~$I_{\text{max}}^{(1)}$, different for all three driving functions~$\Fg(t)$,  has been introduced here to normalize the curves conveniently. 
	}
	}
	\label{fig3}
\end{figure}

In Eq.~(\ref{eq:G1_omega_lowfreq}) the $RC$-time of the system can be identified,
\begin{equation}
	\tauRCo
	\equiv
	\Rtoto\Ctoto=
	\frac{N_\GG+1}{2N_\GG}
	\bigg\{
	\frac{1}{\tau}
	+
	\frac{1}{\tauc}
	+
	\frac{1}{N_\GG\taug}
	\bigg\}^{\!-1}
	,
\end{equation}
which is a relevant time-scale characterizing the response of the system to the gate driving. 
In this section, we focus on analyzing the low-frequency regime, where only the first term of Eq.~(\ref{eq:G1_omega_lowfreq}) needs to be taken into consideration. The applicability of this approximation is, thus,  restricted to frequencies satisfying~\mbox{$\omega\ll1/\tauRCo$}. 
In this regime, the time-resolved current takes the particularly simple
form\footnote{%
We will add the superscript `(1)', whenever necessary, to highlight that a quantity refers to the adiabatic-response regime, where only the term of the first order in the driving frequency contributes.%
} 
\begin{equation}\label{eq:dVgeff1_t_ad}
	\Iad(t)
	=
	D\Ctoto
	\frac{\intd\dVg(t)}{\intd t}
	,
\end{equation}
to which we also refer as the adiabatic-response current.
With respect to Eq.~(\ref{eq:dVgeff1_t}), this means that~$\Ctoto(\omega)\equiv \mbox{$Cg(\omega)\approx\Ctoto$}$, and accordingly,~\mbox{$\dVgeffo(t)\rightarrow\dVg(t)$}, see also Appendix~\ref{app:Self-consist_interact}.  
Equation~(\ref{eq:dVgeff1_t_ad}) clearly demonstrates the characteristic feature of the low-frequency response, namely,  one can essentially obtain the current response of the desired form just by shaping the derivative of a~gate potential~$\dVg(t)=\dVg\Fg(t)$. 
To illustrate this feature, in Fig.~\ref{fig3}(a) we present examples of three different types of the driving function~$\Fg(t)$ together with the resulting current response shown in Fig.~\ref{fig3}(b).

Analyzing the currents in the right contact~$I(t)$, one can immediately notice that both the harmonic and periodic Lorentizan drivings are not suitable for achieving well separated current pulses. On the other hand, the~smooth-box driving allows for emission of a train of positive and negative charge current pulses, with well defined temporal resolution. Since these pulses have a Lorentizan shape, they are expected to carry a quantized amount of charge, when the amplitude is chosen appropriately. This claim is going to be verified in Sec.~\ref{sec:Noise},
\new{
where we calculate the number of excess electrons and holes.
}
\new{%
In order to determine the average number of electrons and holes emitted, care has to be taken whenever these particles are not well separated from each other in time. We define the%
}
average number of electrons~$\avQ$ transferred to the right electrode in a half-period interval as
\begin{equation}\label{eq:avQ}
	\avQ
	=
	D
	\mathcal{Q}
	\Theta_{\text{max}}\ .
\end{equation}
It is given by a product of the transmission probability~$D$ of the QPC,  the number of charges 
\mbox{$
	\mathcal{Q}
	\equiv
	\Ctoto\dVg
	/
	|e|
$}
that can be brought on a ``capacitor'' with capacitance $C_\mu$ when a potential $\delta V_\text{g}$ is applied to it, and the function
\begin{equation}\label{eq:Theta_max}
	\Theta_{\text{max}}
	\equiv
	\max_{\forall0\leqslant t_0<\perT}
	\bigg\{
	\frac{1}{D\Ctoto\dVg}
	\!
	\int\limits_{t_0}^{t_0+\perT/2}
	\!
	\intd t
	\,
	I(t)
	\bigg\}
	.
\end{equation}
In essence, this latter expression describes the process of finding the maximal value of the integral in brackets by scanning the current~$I(t)$,
\new{%
Eq.~(\ref{eq:I1_t_gen}),}
within one period of the driving with a half-period window.
\new{%
It  selects, thus, the appropriate window for electron (and consequently, also hole) emission.%
}
Specifically, it can be checked that in first order in the driving frequency, using Eqs.~(\ref{eq:F_har})-(\ref{eq:F_s-box}), for all three driving signals under consideration
\new{%
in Fig.~\ref{fig3}%
}
 one 
obtains~\new{%
\mbox{$\Theta_{\text{max}}^{(1)}=\Fg(\perT/2)-\Fg(0)=1$}, 
}
and
\begin{equation}\label{eq:avQ_ad}
	\avQad
	=
	D\mathcal{Q}
	.
\end{equation}

To complete the present discussion, we note that the current response~$\Iad(t)$,
\new{
and consequently also~$\avQad$,%
}
is only trivially affected by the choice of the gate-electrode type, quantified by the number of channels~$N_\GG$. We recall that~$N_\GG$ enters the problem through the electrochemical capacitance~$\Ctoto(\omega)$, Eq.~(\ref{eq:Ctot1}),  which in the regime of slow driving becomes frequency-independent, $\Ctoto(\omega)\approx\Ctoto$, so that it acts merely as a scaling factor for the current~(\ref{eq:dVgeff1_t_ad}). Consequently, one expects that the current amplitude for a metallic gate~(\mbox{$N_\GG\rightarrow\infty$}) is \mbox{$1+\Ctoto(N_\GG\rightarrow\infty)/\Cqg$} times larger than the one for a gate employing a single edge channel~(\mbox{$N_\GG=1$}), with \mbox{$\Ctoto(N_\GG\rightarrow\infty)\equiv C\Cqc/\big(C+\Cqc\big)$}.
%

\begin{figure*}[t]
	\includegraphics[width=0.99\textwidth]{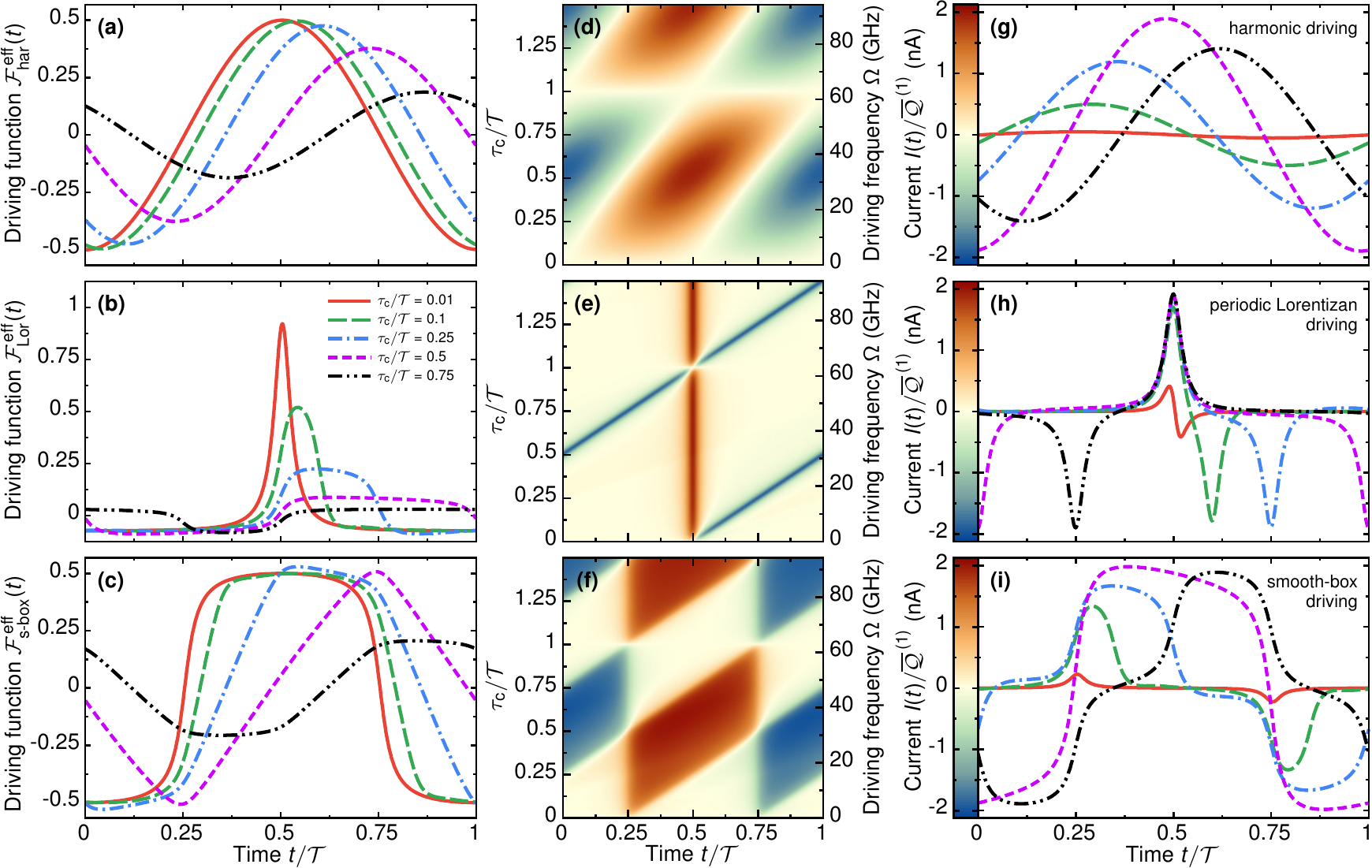}
	\caption{
	The effect of fast gate driving on the time-resolved current response~$I(t)$ in the right reservoir of the conductor, Eq.~(\ref{eq:I1_t_gen}), for the harmonic (a,d,g), Eq.~(\ref{eq:F_har}), periodic Lorentzian (b,e,h), Eq.~(\ref{eq:F_L}), and smooth-box (c,f,i), Eq.~(\ref{eq:F_s-box}), driving functions. 
	\emph{Left column} [(a)-(c)]: evolution of the effective gate potential~\mbox{$\dVgeffo(t)=\dVg\Fgeff(t)$}, Eq.~(\ref{eq:dVgeff1_t}), with increasing~$\tauc/\perT$ \mbox{---see} panel~(b) for the legend.
	\emph{Middle column} [(d)-(f)]: time-resolved map plots of~$I(t)$, scaled by the value in the first order in frequency of the average number of particles~$\avQad$, Eq.~(\ref{eq:avQ_ad}), shown as a function of~$\tauc/\perT$ (see the right $y$-scale for corresponding values of the driving frequency~$\Omega$). 
	\emph{Right column} [(g)-(i)]: cross-section plots of relevant maps in the middle column for selected values of~$\tauc/\perT$ given in~(b). 
	Note that solid lines in~(a)-(c) [(g)-(i)] illustrate the case of the slow driving limit, and they  
	\new{%
	already qualitatively match the adiabatic-response result in Fig.~\ref{fig3}(a)\,[(b)], as expected.%
	}
	Numeric results have been obtained for $\Kmax=50$ harmonics included in the expansion~(\ref{eq:Fg_expand}) for periodic Lorentzian and smooth-box gate drivings, $\ell=10$~$\mu$m ($\tauc=0.1$~ns and $\Cqc\approx3.9$~fF), $C=10$~fF. Additionally, the gate lead has been assumed to support a single edge channel ($N_\GG=1$). 
	For further discussion of parameters used in calculations see Appendix~\ref{app:Self-consist_interact}.
	}
	\label{fig4}
\end{figure*}

\subsection{\label{sec:Non-adiabatic_driving}Arbitrary-frequency driving regime}

The evolution of the time-resolved current response~$I(t)$ when increasing the driving frequency~$\Omega$ for the three selected driving potentials defined in Sec.~\ref{sec:Current_response} is illustrated in Fig.~\ref{fig4}. This figure represents the situation of a gate lead supporting only a single edge channel~(\mbox{$N_\GG=1$}), and we assume the conceptually simplest situation of~$\tauc=\taug$. Later on we will relax both these constraints, and discuss the consequences. 

To begin with, one can see in Fig.~\ref{fig4}(a)-(c) that with the increase of the driving frequency~$\Omega$, which basically translates into a larger and larger ratio~$\tau_{RC}/\mathcal{T}$, the effective gate potential~$\dVgeffo(t)\equiv\dVg\Fgeff(t)$ in the expression for the current~(\ref{eq:I1_t_gen}) is not identical anymore to the one applied to the gate electrode~$\dVg(t)$, as in the adiabatic-response discussed above. Note that in Fig.~\ref{fig4} and following figures, we use as a plotting parameter the quantity~$\tauc/\perT$ instead of $\tau_{RC}/\perT$. 
\new{
This choice is motivated by Eq.~(\ref{eq:Ctot1}) [entering Eq.~(\ref{eq:dVgeff1_t}) for~$\dVgeffo(t)$], which for fixed geometric capacitance $C$ and $\tau_\text{c}=\tau_\text{g}$ is conditioned by~$\tauc/\perT$.%
}
For the parameters chosen in Fig.~\ref{fig4}, $\tau_{RC}$ and $\tauc$ are approximately equal.

The observed difference between $\dVgeffo(t)$ and $\dVg(t)$ occurs because for larger driving frequencies the electrochemical capacitance~$\Ctoto(\omega)$, Eq.~(\ref{eq:Ctot1}), can no longer be regarded as a frequency-independent function. The reason is that in such a high-frequency regime, the quantum-capacitance contributions~$\Cqc(\omega)$ and~$\Cqg(\omega)$ significantly depend on how fast the gate potential is oscillating, see Fig.~\ref{fig1_app} in Appendix~\ref{app:Self-consist_interact}. It means that strongly frequency-dependent screening occurs in this device, similar to what has been observed in Ref.~\cite{Bocquillon_NatureCommun.4/2013}. For large frequencies, the quantum capacitance, Eq.~(\ref{eq:Cqeta_def}), gets suppressed with respect to the geometric capacitance, so that the latter ceases to contribute to the dynamics, meaning that interactions are screened. Remarkably, considerable deviations already occur as soon as the $RC$-time becomes only a fraction of the driving period~$\perT$, clearly showing the importance of the specific driving for the functioning of the device. 

The modification of the effective potential~$\dVgeffo(t)$ with respect to the externally applied gate potential~$\dVg(t)$ affects, in turn, the charge current in the right electrode, as shown in Fig.~\ref{fig4}\mbox{(d)-(i)}. 
Importantly, one can distinguish two characteristic features observed for all three driving functions under consideration, that is, the current signal is quenched for~\mbox{$\tauc/\perT=1$}, whereas for~\mbox{$\tauc/\perT=0.5$} the above discussed screening is strong, so that the 
\new{%
shape of the current response resembles that%
} 
of the actual driving potential in the first half of the period and repeats it with a negative sign in the second half. 
The origin of both these features can be explained by analyzing the behavior of the factor~$\Ctoto(k\Omega)$ in Eq.~(\ref{eq:dVgeff1_t}), which is, in turn, determined by~$\Cqc(k\Omega)$ and~$\Cqg(k\Omega)$, Eq.~(\ref{eq:Cqeta_def}) \mbox{---mind} that in the single-edge case of~\mbox{$\tauc=\taug$}, considered here, \mbox{$\Cqc(k\Omega)=\Cqg(k\Omega)$}. 
One can then immediately conclude that the suppression of the current occurs because for~\mbox{$\tauc/\perT=1$} one finds~$\Cqc(k\Omega)=0$, and consequently, also~$\Ctoto(k\Omega)=0$. 

On the other hand, the screening mechanism behind the effect of reproducing the driving signal by the current is more subtle to understand technically. Let us first reformulate the expression for current~(\ref{eq:I1_t_gen}) 
\new{%
as%
}
\begin{align}\label{eq:current_rewrite}
	I(t)
	\propto
	\sum\limits_{k=1}^{\Kmax}
	\bigg[
	&
	\text{Re}
	\Big\{
	2
	\Fg_k
	\,
	\text{e}^{-ik\Omega t}
	\Big\}
	k\text{Im}
	\big\{\Ctoto(k\Omega)\big\}
\nonumber\\[-2pt]
	+\ &
	\text{Im}
	\Big\{
	2
	\Fg_k
	\,
	\text{e}^{-ik\Omega t}
	\Big\}
	k\text{Re}
	\big\{\Ctoto(k\Omega)\big\}
	\bigg]
	.
\end{align}
One can see that except for the factor~\mbox{$k\text{Im}\{\Ctoto(k\Omega)\}$}, the~first line in the equation above looks much alike the expression~(\ref{eq:Fg_expand}) for the driving function~$\Fg(t)$. It means that this term is expected to dominate in the screened case. To show this, we now concentrate on the ideal case, where~\mbox{$\tauc/\perT=0.5$}. Using Eqs.~(\ref{eq:Cqeta_def}) and (\ref{eq:Ctot1}), we notice that \mbox{$\Cqc(k\Omega)=0$} if~$k$ is \emph{even} [resulting in $C_\mu(k\Omega)=0$] and~\mbox{$\Cqc(k\Omega)=2i\Cqc/(k\pi)$} if~$k$ is \emph{odd}, which for~\mbox{$N_\GG=1$} edge channel in the gate yields
\begin{subequations}
\begin{gather}\label{eq:ReCtoto}
	\text{Re}\big\{\Ctoto(k\Omega)\big\}
	\overset{k\ \mathrm{odd}}{=}
	\frac{
	C\big[2\Cqc\big]^2
	}{
	\raisebox{-1ex}{$\big[2\Cqc\big]^2
	+
	\big[2k\pi C\big]^2$}
	}
	,
\\\label{eq:ImCtoto}
	\text{Im}\big\{\Ctoto(k\Omega)\big\}
	\overset{k\ \mathrm{odd}}{=}
	\frac{
	4k\pi C^2\Cqc
	}{
	\raisebox{-1ex}{$\big[2\Cqc\big]^2
	+
	\big[2k\pi C\big]^2$}
	}
	.
\end{gather}
\end{subequations}
Importantly, in the strongly screened case with \mbox{$C>\Cqc$}, the factor \mbox{$k\text{Re}\big\{\Ctoto(k\Omega)\big\}$} becomes attenuated for higher harmonics as 
\begin{equation}\label{eq:ReImCtoto_Ng_1}
	\frac{
	\text{Re}\big\{\Ctoto(k\Omega)\big\}
	}{
	\raisebox{-0.5ex}{$\text{Im}\big\{\Ctoto(k\Omega)\big\}$}
	}
	=
	\frac{1}{k\pi}
	\cdot
	\frac{\Cqc}{C}
	,
\end{equation}
so that if \mbox{$\Cqc\ll C$}, the second line of Eq.~(\ref{eq:current_rewrite}) is already suppressed with respect to the first one for any value of~$k$. This essentially means that the current~$I(t)$ is in this limit linearly proportional to the odd-order harmonics of the originally applied driving potential, $\dVg^\text{odd}(t)$, so  that   \mbox{$I(t)\approx G\dVg^\text{odd}(t)$} with \mbox{$G\equiv2D\Cqc/\perT$}. See also \mbox{Fig.~\ref{fig3_app}} in Appen\-dix~\ref{app:Aux_figures} for the dependence of the screening on the magnitude of the geometric capacitance $C$.
The~fact that only the component~$\dVg^\text{odd}(t)$ is mapped by the current comes here as a direct consequence of charge conservation within each driving cycle, that is, each negatively charged pulse must be  accompanied within one period by a pulse with the opposite charge. 
This is exactly ensured by the property of an arbitrary periodic function~$f(t)$ consisting of odd harmonics, only:  any feature occurring in such a function at time~$t$ is followed by its inverse-in-sign counterpart at later time~$t+\perT/2$.
The effect under discussion is especially evident for the periodic Lorentzian driving function, see the short-dashed (magenta) line in Fig.~\ref{fig4}(h), which is composed from both even- and odd-order harmonics, unlike the harmonic and the smooth-box functions. 
Consequently, in the 
\new{fast-driving}
limit, the Lorentzian voltage pulse can be used to emit successively single electron and hole wave-packets. This is also reflected by the shape of the effective potential of the gate $\dVgeffo(t)$ which tends to a smooth-box shape for \mbox{$\tauc/\perT\rightarrow0.5$}, as illustrated by the short-dashed (magenta) line in Fig.~\ref{fig4}(b). 
\new{
The Lorentzian drive yields, thus, in the fast-driving limit the same result as the smooth-box drive in the adiabatic-response regime.%
}
Note that in the intermediate driving regime, Lorentzian-shaped current pulses can be obtained only by more intricate design of the \emph{external} driving signal, see Eq.~(\ref{eq:current_rewrite}).

\begin{figure}[t]
	\includegraphics[scale=1.1]{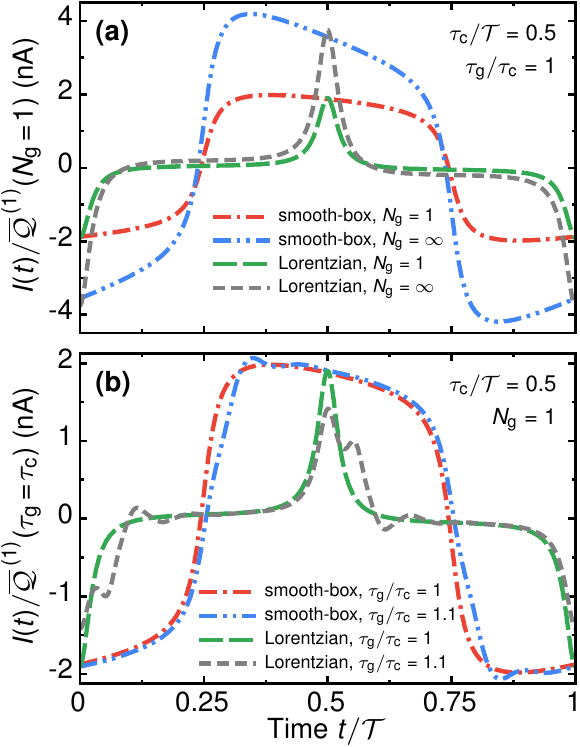}
	\caption{
	Dependence of the time-resolved current~$I(t)$, in the non-adiabatic-response driving regime~(\mbox{$\tauc/\perT=0.5$}), on the type of the gate electrode used, for different driving signals. 	Results obtained for the smooth-box (periodic Lorentzian) driving potential are  indicated by dotted-dashed (dashed) lines. The lines obtained for~\mbox{$N_\GG=1$} correspond to the short-dashed (magenta) lines in Fig.~\ref{fig4}(h)-(i).
 	(a) Comparison between a gate with a single edge channel~(\mbox{$N_\GG=1$}) and a metallic gate~(\mbox{$N_\GG\rightarrow\infty$}).
	(b) Comparison between two gates with single edge states, where $\tau_\text{c}=\tau_\text{g}$ or $\tau_\text{g}/\tau_\text{c}=1.1$. 
	Parameters not indicated are chosen as in Fig.~\ref{fig4}.
	}
	\label{fig5}
\end{figure}

So far we have focused on the case of the gate lead supporting only a single edge channel~(\mbox{$N_\GG=1$}) with \mbox{$\tau_\text{g}=\tau_\text{c}$}. In the following we are going to generalize this situation by addressing 
\new{both%
} 
the case where  \mbox{$\tau_\text{g}\neq\tau_\text{c}$},
\new{
as well as%
}
the metallic case, corresponding to \mbox{$N_\text{g}\rightarrow\infty$}. We start with the discussion of the latter. From Eq.~(\ref{eq:Ctot1}), we expect one trivial consequence, namely, that for \mbox{$N_\GG\rightarrow\infty$} the total capacitance~$\Ctoto$ should increase, leading to an increase of the current~$I(t)$, Eq.~(\ref{eq:I1_t_gen}), as well. 
This amplitude increase is visible in Fig.~\ref{fig5}(a) [see also a corresponding plot of the average number of electrons~$\avQ$, Eqs.~(\ref{eq:avQ})-(\ref{eq:Theta_max}), in Fig.~\ref{fig4_app} in  Appendix~\ref{app:Aux_figures}]. What is more important, however, 
\new{%
is%
} 
that the change of the gate properties has an impact on the screening. This leads to a clear asymmetry in the time-resolved signal with respect to the symmetric situation of the maximally screened case of the setup with a gate supporting a single edge channel.

This less effective screening can be understood by a study analogous to the one for the single-channel gate, leading to the estimate in Eq.~(\ref{eq:ReImCtoto_Ng_1}). For the metallic gate, \mbox{$N_\GG\rightarrow\infty$}, one derives for \mbox{$C\gg\Cqc$} 
\begin{equation}\label{eq:ReImCtoto_Ng_inf}
	\frac{
	\text{Re}\big\{\Ctoto(k\Omega)\big\}
	}{
	\raisebox{-0.5ex}{$\text{Im}\big\{\Ctoto(k\Omega)\big\}$}
	}
	=
	\frac{2}{k\pi}
	\cdot
	\frac{\Cqc}{C}\ .
\end{equation}
This essentially means [cf.~Eqs.(\ref{eq:ReImCtoto_Ng_1}) and~(\ref{eq:ReImCtoto_Ng_inf})] that under the same conditions the system with a metallic gate lead is less efficient 
\new{
(by a factor 2)%
}
in reproducing the shape of the driving potential.

Finally, we come back to the gate supporting a single edge only and address the situation when
\new{
the traversal times~$\tauc$ and~$\taug$ are no longer the same,~\mbox{$\tauc\neq\taug$}. Such a situation can arise, for instance, when the conductor and the gate lead are made of different materials, so that the respective drift velocities vary slightly.
On the other hand, if the same material is used for both the gate and the conductor, slight differences in traversal times can in principle occur due to impurities or rough boundaries of the gate or the conductor, which can effectively increase the length of the interacting region.%
}
As a result, the exactly symmetric behavior between gate and conductor, for example in the total capacitance, Eq.~(\ref{eq:Ctot1}), will not persist and one generally expects new features to arise. For instance, the suppression of the current 
\new{%
previously%
}
observed at $\mathcal{T}=\tau_\text{c}$ 
\new{%
for equal traversal times%
}
is expected to split, that is, it should occur not only if~$\tauc/\perT=1$ but also for~$\taug/\perT=1$, because the condition~\mbox{$\Ctoto(k\Omega)=0$} is now fulfilled either when~\mbox{$\Cqc(k\Omega)=0$} or~\mbox{$\Cqg(k\Omega)=0$}. More generally, new features in the  time-resolved current~$I(t)$ form when~$\tauc\neq\taug$ and the gate potential is driven non-adiabatically. This appearance of new features in the time-resolved current is shown in Fig.~\ref{fig5}(b). See also Appendix~\ref{app:Aux_figures} for a detailed presentation of the impact on the different driving signals in the full frequency regime.
The impact on the current, in particular on the screening effects, resulting from different gate-realizations, clearly shows  the delicate parameter tuning that is required in order to acquire well-separated current pulses carrying quantized charge.

\subsection{\label{sec:Noise}Characteristics of the emitted pulses}

%
Our aim is now to analyze whether the created current pulses indeed represent well-separated signals carrying a quantized charge. From our previous analysis of the time-resolved current, one can already restrict the possible candidates. In the  adiabatic-response regime, the smooth-box potential, Eq.~(\ref{eq:F_s-box}), seems to produce the desired effect [see the solid line in Fig.~\ref{fig3}(b)].  In the non-adiabatic driving regime, on the other hand, the engineering of the suitable gate potential appears to be more challenging. Here, one should rather choose the periodic Lorentzian potential, Eq.~(\ref{eq:F_L}), whose shape  can be recovered in the current signal, for~$\tauc/\perT\approx0.5$ due to strong frequency-dependent screening [see the short-dashed line in Fig.~\ref{fig4}(h)].

\begin{figure}[t]
	\includegraphics[scale=1]{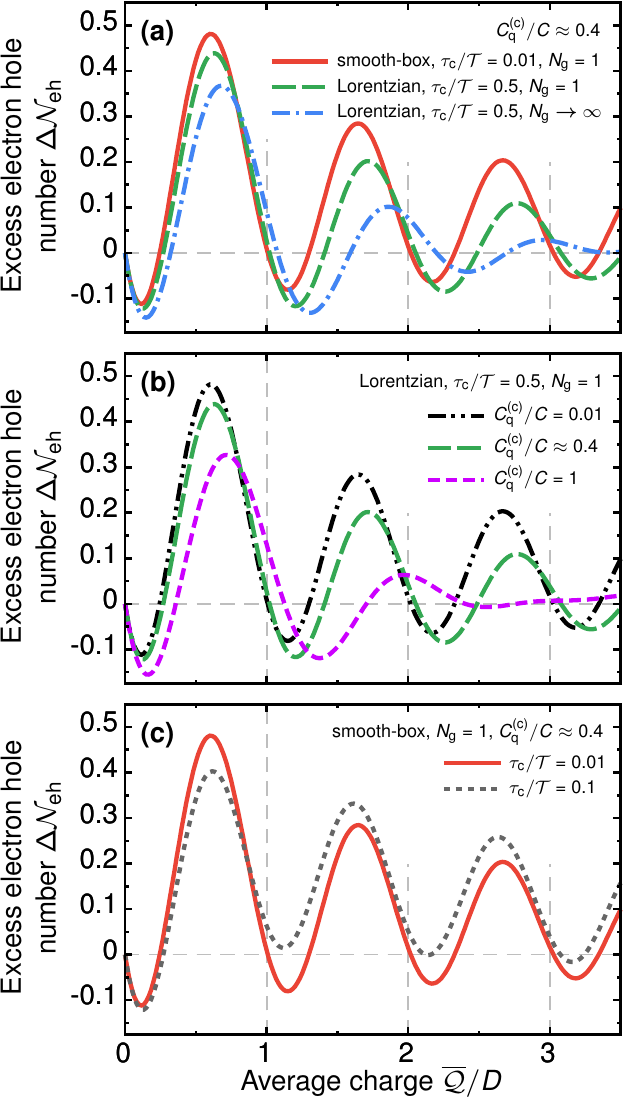}
	\caption{
	Number of excess emitted particles~$\dNeh$~as~a~func\-tion of the  average number of electrons~$\avQ/D$.	%
	(a) Comparison between different types of the gate potential in the  adiabatic-response~(\mbox{$\tauc/\perT=0.01$}) and non-adiabatic~(\mbox{$\tauc/\perT=0.5$}) driving regimes. In the latter regime, the results are shown for a gate supporting a single edge channel~(\mbox{$N_\GG=1$}) and a metallic gate~(\mbox{$N_\GG\rightarrow\infty$}).
	(b) The~effect of the geometric capacitance~$C$ on~$\dNeh$ for the periodic Lorentzian gate potential in the non-adiabatic driving regime~(\mbox{$\tauc/\perT=0.5$}) and~\mbox{$N_\GG=1$}.
	(c) The effect of an increasing frequency for driving with the smooth-box potential. 
	All parameters not indicated are taken as in Fig.~\ref{fig4}.
	}
	\label{fig6}
\end{figure}

While an integration over time of the created current pulses yields the \textit{average} transported charge, no information is obtained about the precision with which a certain amount of charge is injected. In order to acquire this information, namely how many (extra) electron-hole pairs are created by the driving~\cite{Reydellet_Phys.Rev.Lett.90/2003,Vanevic_Phys.Rev.B93/2016,Battista_Phys.Rev.B90/2014,Rychkov_Phys.Rev.B72/2005,Dubois_Nature502/2013}, one needs to consider the zero-frequency current noise,
\new{%
Eq.~(\ref{eq:Noise_def}), for example, in the right contact of the conductor,~$\mathcal{P}_{\RR\RR}\equiv \mathcal{P}$.%
}
In order to keep the following analysis focused, we  
assume~$k_\text{B}T=0$.\footnote{%
\new{%
Note that in the case of the current~$I(t)$, Eq.~(\ref{eq:I1_t_gen}), temperature never plays a role because the scattering in the interacting region does not depend on the incident energy.
}
}
In this limit, the current noise takes the 
form\footnote{%
\new{%
A finite-temperature expression for an arbitrary constant bias drop across the conductor can be found in Appendix~\ref{app:Noise}.
}
}
\begin{equation}
	\mathcal{P}
	=
	\frac{e^2}{\perT}
	D(1-D)
	\!
	\sum_{i=\elec,\hole}
	\!
	\Npart^{i}
\end{equation}
with~$\Npart_1^i$ denoting the number of electron~\mbox{(\mbox{$i=\elec$})} and hole~(\mbox{$i=\hole$}) excitations generated during one driving cycle. This number is defined as
\begin{equation}\label{eq:Npart_1}
	\Npart^{\elec/\hole}
	=
	\sum\limits_{n=1}^{+\infty}
	n
	\big|\Cc_{\pm n}\big|^2
	,
\end{equation}
where the coefficients~$\Cc_{\pm n}$ are explicitly given by Eq.~(\ref{eq:coeffCc}) in Appendix~\ref{app:Deriv_C}.

To analyze the properties of the generated pulses, we introduce the number of excess emitted particles~\cite{Gabelli_Phys.Rev.B87/2013,Dubois_Phys.Rev.B88/2013} defined as
\begin{equation}\label{eq:dNeh}
	\dNeh
	\equiv
	(\Npart^{\elec}+\Npart^{\hole})
	-
	2\avQ/D
	.
\end{equation}
The first term \mbox{$\Npart^{\elec}+\Npart^{\hole}$} counts the \emph{total} number of electrons and holes carried by the signal. The second term~$2\avQ/D$ [with the  average number of electrons~$\avQ$ given in~Eq.~(\ref{eq:avQ})] counts, on the other hand, the number of electrons and holes that one would expect if only the average charge per half-period is counted. In the situation of interest, the charge carried by the pulses is quantized, namely $2\avQ/D=2$. In the ideal case, where
\new{%
on average a quantized charge is emitted per half period and%
}
no extra electron-hole pairs are created, $\dNeh$ equals zero.

\new{%
Note that in contrast to the stationary case~\cite{Gabelli_Phys.Rev.B87/2013,Dubois_Phys.Rev.B88/2013}, where $\dNeh$ is always non-negative, here negative values can occur. The physical reason for this is that \mbox{---also} in the ideal \mbox{case---} charges are always emitted in pairs (one electron and one hole). 
However, when the sought-for electron and hole pulses overlap, there is no unambiguous way to distinguish them from the spurious electron-hole pairs due to the non-ideal operation.
In consequence, when the non-integer number of average charge~$\avQ$ increases faster than the amount of extra electron-hole pairs, $\dNeh$ can become 
negative.\footnote{%
\new{Technically, in the limit of small driving amplitude~$\dVg$  (that is, for \mbox{$|e|\dVg\ll \hbar\Omega$}), the number of electron/hole excitations~$\Npart^{\elec/\hole}$, Eq.~(\ref{eq:Npart_1}), grows at least quadratically in~$\dVg$ [expand Bessel functions in Eq.~(\ref{eq:coeffCc})], while for the current signal one observes \mbox{$I(t)\propto\dVg$}, see Eqs.~(\ref{eq:I1_t_gen})-(\ref{eq:dVgeff1_t}). As a result, \mbox{$\dNeh<0$} at small driving amplitudes inevitably implies the occurrence of an additional (meaningless)  zero.}}
Note that these regimes of~\mbox{$\dNeh\leqslant0$} are not a sign of improvement of the signal. Only in the case where the average charged~$\avQ$ takes an integer value, does the excess particle number have a clear interpretation.
}

In Fig.~\ref{fig6}, we show the number of excess particles  as a function of the  average number of electrons~$\avQ/D$ transferred to the right electrode during one half of the driving cycle, Eq.~(\ref{eq:avQ}). We compare the results for different driving signals and gate geometries. 
As mentioned above, ideal single-particle injection is expected for the adiabatic-response to the smooth-box driving potential and from the Lorentzian-shaped driving in the non-adiabatic regime. In Fig.~\ref{fig6}(a), we thus show the excess particle number~$\dNeh$ in the \emph{adiabatic-response} regime for the smooth-box gate potential (solid line for~\mbox{$\tauc/\perT=0.01$}), and in the \emph{non-adiabatic} regime for the periodic Lorentzian gate potential (long-dashed line for~\mbox{$\tauc/\perT=0.5$}). It can be seen that whereas the smooth-box potential in the slow driving regime yields~$\dNeh\approx0$ for each
integer\footnote{Overlapping of several Lorentzians does not change the 
\new{%
excess%
}
noise at all. It is the energy current and its correlations~\cite{Battista_Phys.Rev.B90/2014}, which would reveal that the single charges are not independent any longer due to the Pauli principle.
}
$\avQ/D$, slight deviations occur for the periodic Lorentzian potential in the non-adiabatic regime, which increase with the amount of ``approximately integer'' charges injected. As discussed in the previous section, the reason for this is that in the latter example for the employed set of parameters the current signal cannot exactly map the shape of the driving potential. This discrepancy becomes even larger when a metallic gate~(dotted-dashed line for~\mbox{$N_\GG\rightarrow\infty$}) is used, which results from a more pronounced asymmetry of the current due to reduced screening, as shown in~Fig.~\ref{fig5}(a).

Next, in Fig.~\ref{fig6}(b), we study the effect of the geometric capacitance~$C$ on the quality of the charge emission for the Lorentzian-shaped driving signal in the non-adiabatic regime. The long-dashed lines in~(a) and~(b) correspond to the same case. The comparison for different values of~$C$ in Fig.~\ref{fig6}(b) shows how an increase of the geometric capacitance, by making the screening more efficient, further improves the emission of quantized charge. As a~result, the solid line in~(a) for the adiabatic-response current to the smooth-box-driving and the double-dotted-dashed line in~(b) perfectly overlap. 
This essentially confirms our previous conclusions drawn from the discussion in which we only considered the shape of the emitted pulses (see Secs.~\ref{sec:Adiabatic_driving}-\ref{sec:Non-adiabatic_driving}). Namely, to obtain the sequence of well-separated pulses carrying a quantized charge, one should choose the smooth-box potential in the adiabatic-response driving regime and the periodic Lorentzian in the non-adiabatic regime.

Finally, panel (c) of Fig.~\ref{fig6} illustrates how sensitively the ideal emission of quantized charge with the smooth-box voltage depends on the smallness of the adiabaticity parameter $\tau_{RC}/\mathcal{T}$. Already for $\tau_\text{c}$ (here of comparable magnitude as $\tau_{RC}$) being an order of magnitude smaller than the driving period, the deviations from the ideal case are substantial.

\section{\label{sec:Conclusions}Conclusions}

The goal of this paper was to investigate the feasibility of creating quantized, noiseless charge pulses by local, time-dependent gating of a conductor in the quantum Hall regime (carrying a single edge state). 
For this purpose, we have carried out a self-consistent study of the time-dependent transport problem which allowed us to analyze how the shape and speed of the gate-voltage driving as well as the geometric and quantum properties of gate and conductor impact the created time-dependent current signal. 
Specifically, we have employed our general theory to three different types of driving potential (harmonic, periodic Lorentzian, and smooth-box) at \emph{arbitrary} driving frequencies. While in the two limiting situations of the adiabatic-response and non-adiabatic driving 
\new{%
for~\mbox{$\tauc\approx\perT/2$}%
}
the design of emission of well-separated  charge excitations is straightforward, it becomes more intricate 
\new{%
for other non-adiabatic driving-frequencies.%
}

We have found that in the \emph{adiabatic-response} regime of slow driving, the injected current signal is directly proportional to the derivative of the applied potential. In this regime the smooth-box driving potential \mbox{---yielding} Lorentzian-shaped \mbox{derivatives---} leads to well-separated pulses of opposite charge. Moreover, we have demonstrated that no excess noise is created at the QPC on which the signal in our setup impinges, as long as the amplitude of the driving is adjusted in such a way that each pulse carries an integer multiple of the electron charge.
This suppression of excess noise is the evidence of true single-particle emission (in the absence of extra electron-hole pair creation due to the driving).

On the contrary, in the \emph{non-adiabatic} driving regime, we have found that high driving frequencies result in strong screening. In consequence, the shape of the created current signal can follow approximately the shape of the driving potential (repeated by the same signal with an opposite sign). Importantly, this screening is particularly effective when the gate has the same properties as the conductor (namely, having a single-edge state as well) and when the geometric capacitance of the setup is large with respect to the quantum capacitance. In such a regime, modulation of the gate with a~Lorentzian-shaped driving potential generates an equally ``clean'' single-particle emission as the smooth-box potential in the adiabatic-response regime.

Our detailed analysis provides insight about the subtle dependence of the created current pulses on the choice of the different parameter regimes, setting out at the same time strategies for the optimal current-pulse generation. Interestingly, such a generation of quantized current pulses by local gating opens up routes for the investigation of single-particle physics and quantum optics with electrons also in systems and materials, in which confinement by QPCs (which would be required for creation of quantum-dot-like single-electron devices) remains  challenging so far.

\acknowledgements

We thank Erwann Bocquillon for careful reading of the manuscript.
Financial support from the Knut and Alice Wallenberg Foundation (J.\,S. and M.\,M.) and from the ERC consolidator grant `EQuO' No.~648236 (G.\,F.) is acknowledged. M.\,M. also acknowledges financial support from the Polish Ministry of Science and Education through a young scientist fellowship (0066/E-336/9/2014).

\appendix

\section{\label{app:Self-consist_interact}\\Self-consistent derivation of the internal potentials}

The aim of this appendix is to determine the internal potentials~$\dUc(t)$ and~$\dUg(t)$ self-consistently, following previous work presented in Refs.~\cite{Buttiker_Phys.Rev.Lett.70/1993,Buttiker_IlNuovoCimentoB110/1995,Buttiker_NATOASI326/1996,Buttiker_NATOASI345/1997}.
In general, one can treat electronic interactions in scattering problems self-consistently as long as backscattering in the interacting region is absent. In the case  treated in this paper, the response of a one-dimensional electron system to externally applied potentials is intrinsically linear~\cite{Cuniberti_Phys.Rev.B57/1998,Mora_Nat.Phys.6/2010,Litinski_Phys.Rev.B96/2017}.

To begin with, employing the scattering matrices~(\ref{eq:SFnc_A_def})-(\ref{eq:SFng_A_def}), we calculate currents $I_{\alpha}(t)$ in all contacts [i.e., for $\alpha=\LL(\text{eft}),\RR(\text{ight}),\GG(\text{ate})$], see~Eqs.~(\ref{eq:opI_def})-(\ref{eq:It_def}),
\begin{gather}\label{eq:IL_t_A}
	I_{\LL}(t)
	=
	g_0
	D
	\big[V_\RR-V_\LL\big]
	+
	g_0
	\big[1-D\big]
	\Delta\dUc(t)
	,
\\[5pt]\label{eq:IR_t_A}
	I_{\RR}(t)
	=
	g_0
	D
	\big[V_\LL-V_\RR
	+
	\Delta\dUc(t)
	\big]
	,
\end{gather}
taking the internal 
potential~\mbox{$\Delta\dUeta(t)
	=
	\dUeta(t)-$}
	\mbox{$\dUeta(t-\tau_\eta)$}
as input. Equally, one can write for the gate
\begin{equation}\label{eq:Ig_t_A}
	I_{\GG}(t)
	=
	N_\GG g_0
	\big[
	\Delta\dUg(t)
	-
	\Delta\dVg(t)
	\big]
	.
\end{equation}
As mentioned above, we now see that the currents~(\ref{eq:IL_t_A})-(\ref{eq:Ig_t_A}) are linear with respect to both the externally applied~[$V_\alpha$] and internal~[$\dUeta(t)$] potentials, which essentially stems from the fact that the conductor and the gate are assumed to be intrinsically ballistic and no backscattering occurs in the potential region affected by the time-dependent driving. In order to obtain these equations from the general equation in terms of Floquet scattering matrices, introduced in Sec.~\ref{sec:Model}, we have made use of some summation rules deriving from the unitarity of the Floquet scattering matrix~\cite{Moskalets_Phys.Rev.B66/2002,Moskalets_Phys.Rev.B69/2004,Moskalets_book}.  For~$\mathcal{C}^{(\eta)}_n$ the following relation can be proven to hold
\begin{multline}\label{eq:Sum_rule_Ceta}
	\sum\limits_{n,m=-\infty}^{+\infty}
	\!\!
	(n\hbar\Omega)^p
	\,
	\mathcal{C}_n^{(\eta)\ast} 
	\mathcal{C}_{n+m}^{(\eta)}
	\,
	\text{e}^{-im\Omega t}
\\
	=
	\delta_{p0}
	+
	\delta_{p1}
	e
	\Delta \dUeta(t)
	.
\end{multline}
Worthy of note is that the analogous formula is valid also for coefficients~$\cVg_n$, Eq.~(\ref{eq:cVg}), that is, with the second term in the right-hand side of Eq.~(\ref{eq:Sum_rule_Ceta}) being substituted with~$\delta_{p1}e\Delta\dVg(t)$.

Next, we assume that a surface~$\Pi$ exists, as depicted in Fig.~\ref{fig2}(b) by the long-dashed line, which encloses the parts of the system that interact electrostatically with each other in such a way that no electric field lines penetrate this surface. This, in turn, means that the pile-up charge 
\new{%
[here  referred to as $\dQeta(t)$]%
}
both in the conductor ($\eta=\CC$) and in the gate plate ($\eta=\GG$) can be related to the internal potentials $\dUeta(t)$ through the geometric capacitance $C$ as
\begin{equation}
	-
	\dQeta(t)
	=
	C
	\big[\dUeta(t)-\dUbareta(t)\big]
	,
\end{equation}
with the notation $\overline{\eta}$ to be understood as~$\overline{\CC}=\GG$ and~$\overline{\GG}=\CC$. 
%
Since the charge in each conductor must be conserved, the temporal accumulation and depletion of charge in the interacting region is accompanied by a flow of charge in the conductors through the surface~$\Pi$, which manifests itself as the continuity equations\footnote{Recall that the positive current $I_\alpha(t)$ corresponds to electrons flowing into the direction of the reservoir~$\alpha$, or conversely, to a~\emph{positive} charge flowing out of the reservoir $\alpha$.}
\begin{gather}\label{eq:Q_conserv_1}
	-\frac{\intd \delta Q^{(\CC)}(t)}{\intd t}
	=
	I_{\LL}(t)+I_{\RR}(t)
	,
\\\label{eq:Q_conserv_2}
	-\frac{\intd \delta Q^{(\GG)}(t)}{\intd t}
	=
	I_{\GG}(t)
	.
\end{gather}

The solution for the internal potentials~$\dUeta(t)$ can conveniently be achieved by considering the problem in frequency space. Then, Eqs.~(\ref{eq:IL_t_A})-(\ref{eq:Ig_t_A}) take the form
\begin{gather}\label{eq:IL_omega_A}
\begin{align}
	I_{\LL}(\omega)
	=\ &
	g_0
	D
	\big[V_\RR(\omega)-V_\LL(\omega)\big]
\nonumber\\
	&-
	i\omega
	\Cqc(\omega)
	\big[1-D\big]
	\dUc(\omega)
	,
\end{align}
\\[5pt]\label{eq:IR_omega_A}
\begin{align}
	I_{\RR}(\omega)
	=\ &
	g_0
	D
	\big[V_\LL(\omega)-V_\RR(\omega)\big]
\nonumber\\
	&-
	i\omega
	\Cqc(\omega)
	D
	\dUc(\omega)
	,
\end{align}
\\[5pt]\label{eq:Ig_omega_A}
	I_{\GG}(\omega)
	=
	-
	i\omega
	N_\GG
	\Cqg(\omega)
	\big[
	\dUg(\omega)
	-
	\dVg(\omega)
	\big]
	.
\end{gather}
with~\mbox{$V_{\LL/\RR}(\omega)=2\pi V_{\LL/\RR}\delta(\omega)$}, since we are interested in constant (and eventually equal) potentials applied to the conductor contacts.
Moreover, in the equations above we have introduced some auxiliary frequency-dependent capacitances defined as
\begin{equation}\label{eq:Cq_omega_app}
	\Cqeta(\omega)
	=
	\Cqeta
	\chieta(\omega)
	,
\end{equation}
where~$C_\text{q}^{(\eta)}=g_0\taueta$ stands for the quantum capacitance~\cite{Buttiker_Phys.Lett.A180/1993,Luryi_Appl.Phys.Lett.52/1998,Gabelli_Science313/2006}, being essentially related to the local density of states in a given channel supporting transport of electrons,  and
\begin{equation}\label{eq:chi_omega_app}
	\chieta(\omega)
	=
	i\,
	\frac{
	1-\text{e}^{i\omega\taueta}
	}{
	\omega\taueta
	}
	.
\end{equation}
\begin{figure}[t]
	\includegraphics[scale=1]{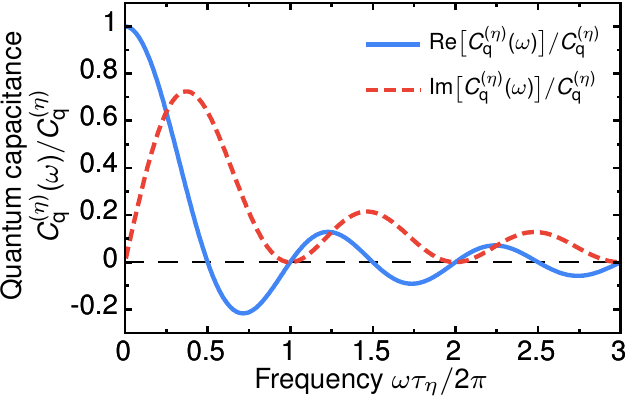}
	\caption{
	Dependence of the real and imaginary part of the quantum capacitance function~$\Cqeta(\omega)$, Eq.~(\ref{eq:Cq_omega_app}), on frequency~$\omega$ scaled to the dwell time~$\taueta$.
	The thin dashed line at 0 serves as a guide for the eye.
	}
	\label{fig1_app}
\end{figure}

In general, the quantum capacitance~$\Cqeta(\omega)$ displays an attenuated oscillatory behavior as a function of frequency~$\omega$, with the period of these oscillations determined by the dwell time~$\taueta$, see Fig.~\ref{fig1_app}. For sufficiently small frequencies~\mbox{$\omega\ll2\pi/\taueta$}, one finds  \mbox{$\Cqeta(\omega)\approx\Cqeta$}, with a vanishingly small imaginary part. In the opposite limit of large~$\omega$ one observes~\mbox{$|\Cqeta(\omega)|\rightarrow0$}, and \mbox{$\Cqeta(\omega)=0$} for all frequencies which are non-zero integer multiples of the characteristic frequency~$2\pi/\taueta$.

Using Eqs.~(\ref{eq:IL_omega_A})-(\ref{eq:Ig_omega_A}) together with the Fourier-transform of Eqs.~(\ref{eq:Q_conserv_1})-(\ref{eq:Q_conserv_2}), one finds
\begin{equation}\label{eq:dUeta_omega_A_app}
	\dUeta(\omega)
	=
	\alphaeta(\omega)
	\,
	\dVg(\omega)
	,
\end{equation}
with the lever arm functions~$\alphaeta(\omega)$ for the conductor ($\eta=\CC$) and the gate ($\eta=\GG$) given by
\begin{gather}\label{eq:alphac_omega_A_app}
	\alphac(\omega)
	=
	\frac{\Ctoto(\omega)}{\Cqc(\omega)}
	,
\\[5pt]\label{eq:alphag_omega_A_app}
	\alphag(\omega)
	=
	\bigg[
	\frac{1}{\raisebox{-0.5ex}{$C$}}+\frac{1}{\Cqc(\omega)}
	\bigg]
	\Ctoto(\omega)
	.
\end{gather}
Above, $\Ctoto(\omega)$ denotes the total (electrochemical) capacitance of a purely electrostatic capacitance~$C$ connected in series with quantum capacitances~$\Cqc(\omega)$ and~$N_\GG\Cqg(\omega)$, that is,
$
	1/\Ctoto(\omega)
	=
	1/C
	\!+\!
	1/\Cqc(\omega)
	\!+\!
	1/\big[N_\GG\Cqg(\omega)\big]
	,
$ see Eq.~(\ref{eq:Ctot1}) in the main text.
Here, the term $N_\GG\Cqg(\omega)$ represents the effective capacitance of a system of quantum capacitors connected in parallel  associated with~$N_\GG$ edge channels in the gate plate.
Noteworthily, it can be checked that after inserting the potentials~(\ref{eq:dUeta_omega_A_app}) into  Eqs.~(\ref{eq:IL_omega_A})-(\ref{eq:Ig_omega_A}), we obtain the currents $I_\alpha(\omega)=\sum_{\beta=\LL,\RR,\GG}G_{\alpha\beta}(\omega)\delta V_{\beta}(\omega)$ for $\alpha$=\LL,\RR,\GG, with the admittance~$G_{\alpha\beta}(\omega)$, which satisfies current conservation, $\sum_\alpha G_{\alpha\beta}(\omega)=0$, and is invariant to a global potential shift, $\sum_\beta G_{\alpha\beta}(\omega)=0$.

Equations~(\ref{eq:dUeta_omega_A_app})-(\ref{eq:alphag_omega_A_app}) fully determine the self-consistent potentials in frequency space.~The general expression for the time-dependent internal potentials~$\dUeta(t)$ is obtained  by performing the inverse Fourier transform of Eq.~(\ref{eq:dUeta_omega_A_app}) for the AC gate potential~$\dVg(t)$,
\begin{equation}\label{eq:dUeta_t_A_gen_app}
	\dUeta(t)
	=	
	\int\limits_{-\infty}^{+\infty}
	\!\!
	\intd t^\prime
	\,
	\alphaeta(t-t^\prime)
	\dVg(t^\prime)
	.
\end{equation}
Furthermore, if one employs the expansion of~$\dVg(\omega)$ into harmonics, see Eqs.~(\ref{eq:Fg_expand}) and~(\ref{eq:Fg_expand_FT}), one finds
\begin{equation}\label{eq:dUeta_t_A_app}
	\dUeta(t)
	=
	\dVg
	\sum\limits_{k=1}^{\Kmax}
	\text{Re}
	\Big\{
	2\Fg_k
	\,
	\text{e}^{-ik\Omega t}
	\,
	\alphaeta(k\Omega)
	\Big\}
	.
\end{equation}
Importantly, analysis of Eqs.~(\ref{eq:dUeta_t_A_gen_app})-(\ref{eq:dUeta_t_A_app}) leads to the observation that the time evolution of the internal potential~$\dUeta(t)$ is governed by the competition between the time scale imposed by the driving, \mbox{$\perT=2\pi/\Omega$}, and  the time scales inherently associated with the system. In particular, these characteristic times are the dwell times~$\tauc$ and~$\taug$, as well as the time scale~\mbox{$\tau=C/g_0$} related to a capacitive coupling between the conductor and the gate. 
To illustrate this aspect, let us discuss how the lever arm~$\alphac(\omega)$ is influenced by the change of the ratio~$\tauc/\perT$. For this purpose, we assume that electrons propagate in the edge channel with the drift velocity~$\vd\approx10^5$~m/s~\cite{Sukhodub_Phys.Rev.Lett.93/2004,McClure_Phys.Rev.Lett.103/2009,Kumada_Phys.Rev.B84/2011}, for simplicity taken the same both in the conductor and the gate, and only \mbox{$N_\GG=1$} transport channel in the gate plate contributes. Moreover, we assume that the gate is driven at a constant frequency~\mbox{$\Omega=10$}~GHz (\mbox{$\perT\approx0.6$}~ns), so that various values of~$\tauc/\perT$  correspond in fact to different lengths~$\ell$ of the interacting region. We furthermore take \mbox{$\tauc=\taug$}. The geometric capacitance~$C=10$~fF (\mbox{$\tau\approx0.26$}~ns)~\cite{Feve_PhysicaE76/2016} is kept constant in the calculations. To keep the discussion simple, we disregard a possible length-dependence of~$C$.

\begin{figure}[t]
	\includegraphics[scale=1]{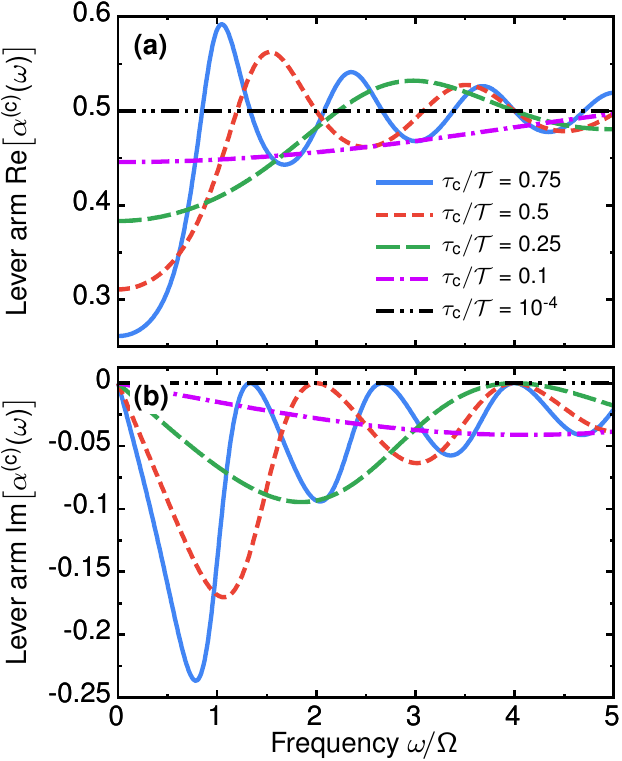}
	\caption{
	Real (a) and imaginary (b) part of the lever arm~$\alphac(\omega)$, Eq.~(\ref{eq:alphac_omega_A_app}), plotted as a function of frequency~$\omega$ scaled to the frequency~$\Omega$ of the gate driving  for indicated values of~$\tauc/\perT$.
	Note that the double-dotted-dashed line in~(a)-(b) corresponds to the limit of the slow driving (\mbox{$\tauc\ll\perT$}), in which $\alphac(\omega)$ can be treated as frequency-independent, that is, \mbox{$\alphac(\omega)\approx\alphac(\omega=0)$}.
	\new{%
	Although $\alphac(\omega)$ is plotted here for a continuous range of~$\omega$, only its values at integer multiples of~$\Omega$ will actually enter the problem, as can be seen in Eq.~(\ref{eq:dUeta_t_A_app}).%
	}
	For discussion of parameters used in calculations see the main text.
	}
	\label{fig2_app}
\end{figure}

The dependence of the lever arm~$\alphac(\omega)$ on the frequency~$\Omega$ of the driving ---which determines the time evolution of~$\dUeta(t)$, as can be seen in Eq.~(\ref{eq:dUeta_t_A_app})--- is shown in Fig.~\ref{fig2_app} for indicated values of~$\tauc/\perT$. As explained above, these values of~$\tauc/\perT$ translate into the length~$\ell$, here ranging from~\mbox{$\approx6$}~$\mu$m for \mbox{$\tauc/\perT=0.1$} (dotted-dashed lines) to~\mbox{$\approx47$}~$\mu$m for \mbox{$\tauc/\perT=0.75$} (solid lines), with~\mbox{$\tauc/\perT=10^{-4}$} 
\new{%
(\mbox{$\ell\approx6$}~nm, double-dotted-dashed lines)%
}
to be understood as the limiting case of pure adiabatic response.
The decrease of the dwell time~$\tauc$ with respect to the period~$\perT$ essentially means that from the point of view of electrons traversing the interacting region, the change of the internal potential gets slower and slower. 
As a result, the ratio~$\taueta/\perT$ (for \mbox{$\taueta=\text{max}\{\tauc,\taug\}$}) can be effectively seen as the adiabaticity parameter describing how fast the system is driven, with~\mbox{$\taueta/\perT\ll1$} corresponding to the \emph{adiabatic-response} regime. It can be seen that when this  limit is approached, the lever arm~$\alphaeta(\omega)$ becomes frequency-independent on the time scale set by the gate driving, that is, \mbox{$\alphaeta(\omega)\approx\alphaeta(\omega=0)\equiv\alpha^{\eta(1)}$}, see the double-dotted-dashed lines in Figs.~\ref{fig2_app}(a)-(b). Note that this reasoning also applies to the capacitances~$\Ctoto(\omega)$ and~$\Cqeta(\omega)$. Equation~(\ref{eq:dUeta_t_A_app}) then reduces to the simple form,
\begin{equation}
	\dUeta(t)
	\rightarrow
	\alpha^{\eta(1)}
	\,
	\dVg(t)
	,
\end{equation}
which represents nothing but the fact that in the adiabatic-response regime the time evolution of the internal potential~$\dUeta(t)$ exactly follows that of the driving potential~$\dVg(t)$.
Finally, we note that the same conclusion  can be reached if one assumes that the length~$\ell$ of the interacting region is kept constant, while one changes the driving frequency~$\Omega$.

\section{\label{app:Aux_fun}\\The periodic Lorentzian function:\\ Auxiliary formulae}

In this appendix, we collect useful formulas for the analytical and numerical treatment of currents emerging from time-dependent driving with a Lorentzian time profile. A periodic driving function $\fLor(t)$ representing a train of Lorentzian pulses with equal amplitudes has the general form
\begin{equation}\label{eq:Lorentzian_def_app}
	\fLor(t)
	=
	\frac{1}{\pi}
	\sum\limits_{n=-\infty}^{+\infty}
	\sum\limits_{k=1}^{N_\text{p}}
	\frac{\perT\Gamma}{
	\big[t-\tpk-n\perT\big]^2
	+\Gamma^2
	}
	,
\end{equation}
where $\Gamma$ stands for the half width at half maximum, and~$\tpk$ describes the position of the~$k$th~Lorentzian pulse (out of a total number of~$N_\text{p}$ pulses) within a single period $\perT$, that is, \mbox{$0\leqslant \tpk\leqslant\perT$}. In the following, we limit our discussion to the case of a single pulse, $\tpk=\delta_{k1}\tp$, as the obtained results can be straightforwardly generalized to the multi-pulse case. For practical reasons, both for analytical and numerical calculations, it is convenient to reformulate Eq.~(\ref{eq:Lorentzian_def_app}) so that the summations are eliminated. 
To do so, let us first rewrite Eq.~(\ref{eq:Lorentzian_def_app}) in a product form,
\begin{multline}
	\fLor(t)
	=
	\frac{\Gamma}{\pi}
	\sum\limits_{n=-\infty}^{+\infty}
	\frac{1}{
	\raisebox{-1.5ex}{
	$n+\dfrac{t-\tp+i\Gamma}{\perT}$
	}
	}
\\
	\times 
	\frac{1}{
	\raisebox{-1.5ex}{
	$n+\dfrac{t-\tp-i\Gamma}{\perT}$
	}
	}	
	.
\end{multline}
Employing the series representation of the digamma function~$\psi(t)$~\cite{Nikiforov_book},
\begin{equation}
	\psi(t)
	=
	-\gamma
	+
	\sum\limits_{n=0}^{+\infty}
	\bigg\{
	\frac{1}{n+1}
	-
	\frac{1}{n+t}
	\bigg\}
	,
\end{equation}
with $\gamma$ denoting the Euler-Mascheroni constant, and the reflection  
formula~$
	\psi(t)-\psi(1-t)
	=
	-\pi\cot(\pi z)
$~\cite{Abramowitz_book},
one obtains
\begin{align}
	\fLor(t)
	=\ &
	\frac{i}{2}
	\sum\limits_{\lambda=\pm1}
	\!
	\lambda
	\cot\!\big(\Omega[t-\tp+i\lambda\Gamma]/2\big)
\nonumber\\
	=\ &
	\frac{
	\sinh\!\big(\Omega\Gamma\big)
	}{
	\cosh\!\big(\Omega\Gamma\big)
	-
	\cos\!\big(\Omega[t-\tp]\big)
	}
	.\label{eq_Lor_product}
\end{align}
This simple expression was for instance used for plots of the pure driving functions as shown in Fig.~\ref{fig3}.

For the derivation of the smooth-box potential, but also in order to calculate average charges produced by the Lorentzian driving, one needs the integral over Eq.~(\ref{eq_Lor_product}). This can be found to be
\begin{equation}
	\int
	\!
	\intd t
	\,
	\fLor(t)
	=
	\frac{i}{\Omega}
	\,
	\text{ln}
	\bigg\{
	\!\!
	-
	\frac{
	\sin\!\big(\Omega[t-\tp+i\Gamma]/2\big)
	}{
	\sin\!\big(\Omega[t-\tp-i\Gamma]/2\big)
	}
	\bigg\}
	,
\end{equation}
which should be understood as a principal value solution, that is, with the imaginary part of $\text{ln}(z)$ lying in the interval $(-\pi,\pi]$.
Moreover, from the equation above it can be concluded that within one period~$\perT$ the area under the driving function~$\fLor(t)$ is unitary,  
\mbox{
$
	(1/\perT)
	\int_0^\perT
	\!
	\intd t
	\,
	\fLor(t)
	=
	1
	.
$
}
On the other hand, the derivative of~$\fLor(t)$ has the form
\begin{align}
	\frac{\intd\fLor(t)}{\intd t}
	=\ &
	\frac{\Omega}{2}
	\,
	\text{Im}
	\Big\{
	\sin^{-2}\!\big(\Omega[t-\tp+i\Gamma]/2\big)
	\Big\}
\nonumber\\
	=\ &
	-
	\frac{
	\Omega
	\sinh\!\big(\Omega\Gamma\big)
	\sin\!\big(\Omega[t-\tp]\big)
	}{
	\big\{
	\cosh\!\big(\Omega\Gamma\big)
	-
	\cos\!\big(\Omega[t-\tp]\big)
	\big\}^2
	}
	.
\end{align}
This formula is helpful, whenever the adiabatic response currents to a Lorentzian driving are evaluated.

\section{\label{app:Aux_figures}Effect of screening -- additional figures} 

\begin{figure}[t]
	\includegraphics[scale=1]{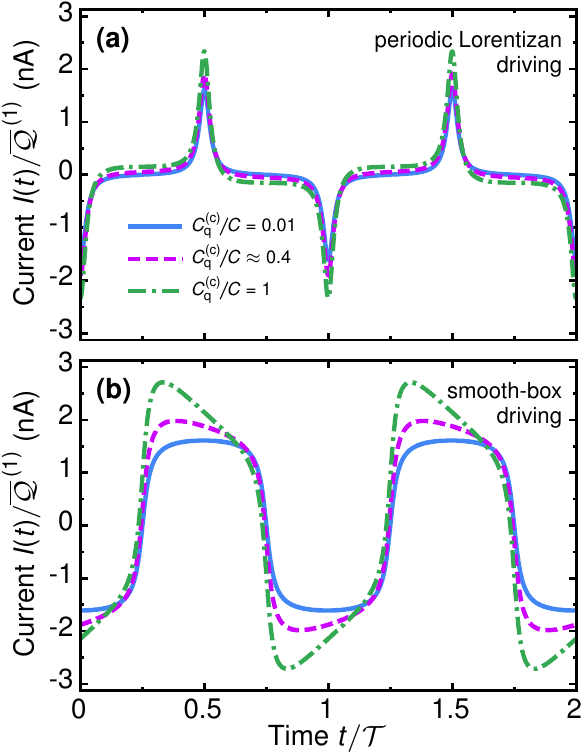}
	\caption{
	Influence of the geometric capacitance~$C$ on the current response~$I(t)$ in the regime of non-adiabatic gate driving for~$\tauc/\perT=0.5$ in the case of the periodic Lorentzian~(a) and smooth-box~(b) functions. 
	Note that the dashed lines presented here are identical to the lines of the same pattern plotted in Fig.~\ref{fig4}(h)-(i).
	Other parameters as in Fig.~\ref{fig4}.
	}
	\label{fig3_app}
\end{figure}

In this appendix, we provide additional material for the study of screening. As discussed in Sec.~\ref{sec:Non-adiabatic_driving}, for~$\tauc/\perT=0.5$ the current signal~$I(t)$ follows approximately the shape of the gate driving potential~$\dVg(t)$, with a copy of an opposite sign in the second half period. This effect becomes particularly visible if the geometrical capacitance~$C$, which stems from electrostatic interaction between the conductor and the gate, is significantly larger than the magnitude of the quantum capacitances~\mbox{$\Cqc=\Cqg$} associated with the time during which an electron is subject to the internal potential. This aspect is illustrated in Fig.~\ref{fig3_app}.

\begin{figure}[t]
	\includegraphics[scale=1]{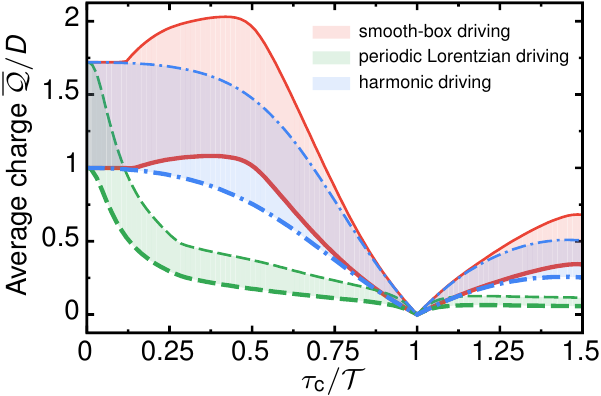}
	\caption{
	The  average number of electrons~$\avQ/D$ transferred to the right electrode in a half-period, Eqs.~(\ref{eq:avQ})-(\ref{eq:Theta_max}). 
	Shaded areas represent achievable values of~$\avQ/D$ when changing the type of the gate electrode between the two limiting cases: one for a single edge channel~(\mbox{$N_\GG=1$}, bold lines), and the other for a metallic gate~(\mbox{$N_\GG\rightarrow\infty$}, thin lines).
	Specifically, the three different types of curves illustrate the results obtained for the harmonic~(dotted-dashed lines), periodic Lorentzian~(dashed lines) and smooth-box~(solid lines) driving potentials, with the bold lines corresponding to the middle column of Fig.~\ref{fig4}.
	}
	\label{fig4_app}
\end{figure}

\begin{figure*}[t]
	\includegraphics[width=0.99\textwidth]{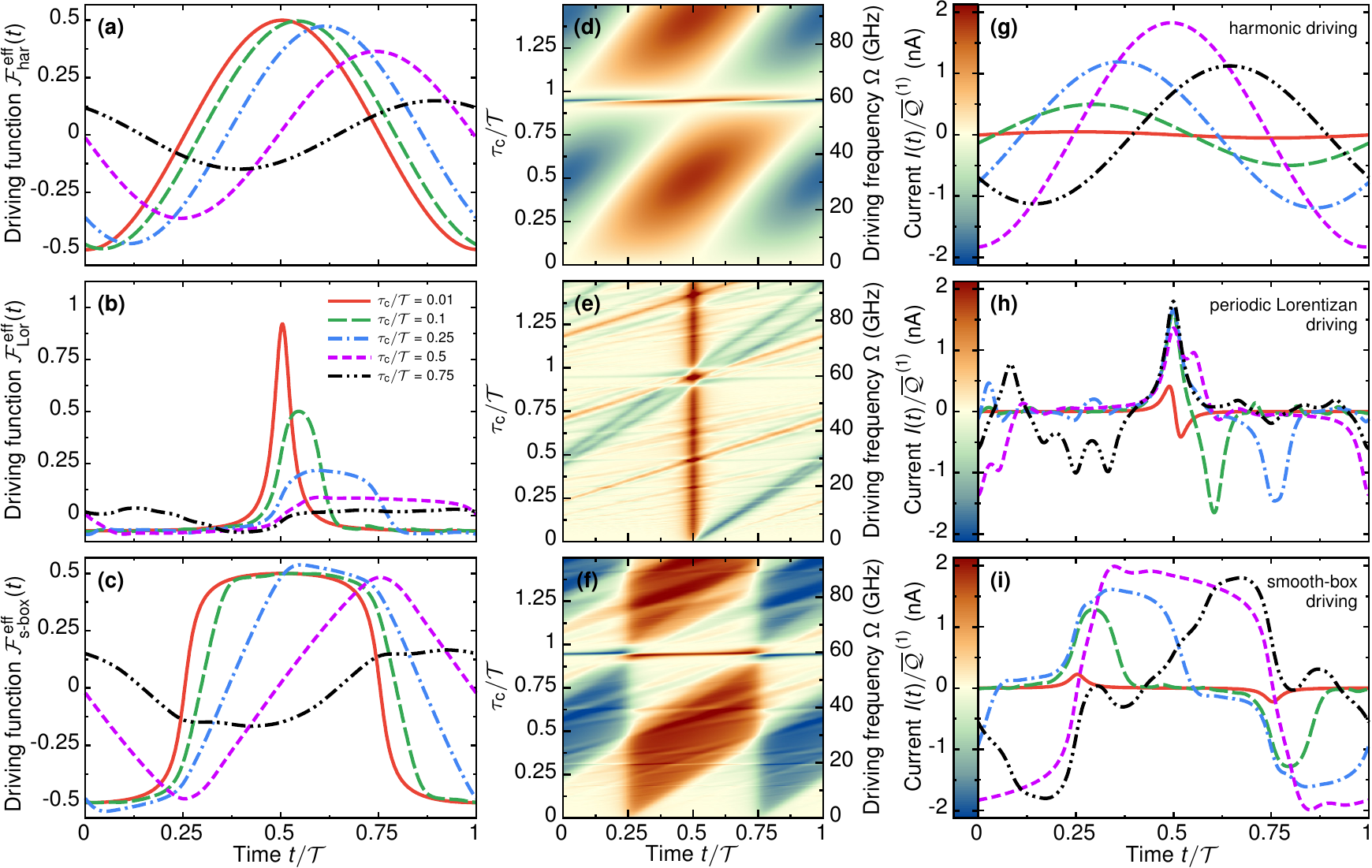}
	\caption{
	Analogous to Fig.~\ref{fig4} except that now the case of~$\taug=1.1\tauc$ is shown.
	All other parameters as in Fig.~\ref{fig4}.
	}
	\label{fig5_app}
\end{figure*}

In fact, we recall that already when studying differences in the gate characteristics in Sec.~\ref{sec:Non-adiabatic_driving} we observed that the effect of screening, appearing in the regime of fast driving, is sensitive to changes in the capacitive properties. 
We discussed there the difference between a metallic gate (\mbox{$N_\text{g}\rightarrow\infty$}) and a gate supporting a single edge channel (\mbox{$N_\text{g}=1$}).
Here,  the (trivial effect of) differences in the current amplitude, as well as the specific, complex differences occurring for different driving signals, are shown in Fig.~\ref{fig4_app}, where the maximum average charge per half-period~$\avQ$, Eq.~(\ref{eq:avQ}), is plotted as a function of $\tau_\text{c}/\perT$.
The variation in~$\avQ$ between a gate supporting just a single edge channel (bold lines) and a metallic gate (thin lines) turns out to depend strongly on the type of driving function, and it is the smallest for the periodic Lorenztian driving function (dashed lines).
In particular, the local maximum in~$\avQ$ developing when approaching~\mbox{$\tauc/\perT=0.5$} for the smooth-box driving function (solid lines) occurs to be especially sensitive to the screening properties.

Finally, an important factor that influences the time-resolved non-adiabatic current response of the system is related to the difference between the traversal times~$\tauc$ and~$\taug$. This can basically be understood as a difference in the quantum capacitance~\mbox{$\Cqc(k\Omega)\neq\Cqg(k\Omega)$}.
The consequences arising from this difference, were discussed in Sec.~\ref{sec:Non-adiabatic_driving} on the example of the periodic Lorentzian driving potential, see Fig.~\ref{fig5}(b). 
To gain more insight into how the results discussed in Sec.~\ref{sec:Non-adiabatic_driving} become modified once~\mbox{$\tauc\neq\taug$}, we show the whole map of effective potentials and resulting current responses in Fig.~\ref{fig5_app}. These results, obtained for~$\taug=1.1\tauc$, are analogous to what has been discussed for the ideal, symmetric case in Fig.~\ref{fig4} in the main text. This overall map shows that additional features occur for all driving signals if~\mbox{$\tauc\neq\taug$}. The effect is particularly strong, when the driving signal contains a large number of harmonics.
The additional features can have a damaging effect on the separation in time (space) of the distinct pulses. Also, deviations from the Lorentzian behavior are expected to lead to an increase of electron-hole pair production.

\section{\label{app:Noise}Zero-frequency charge-current noise}

In the main text, we have limited the discussion of the charge current noise to the zero-temperature regime. Here, we present the general expression for the finite-temperature noise, that is, when \mbox{$k_\text{B}T\neq0$}. Using the definition of the zero-frequency charge-current noise~(\ref{eq:Noise_def}) together with the Floquet scattering matrix~(\ref{eq:SFnc_A_def}), one finds the noise detected in the right contact~\mbox{$\mathcal{P}_{\RR\RR}\equiv \mathcal{P}$} to have the following form:
\begin{align}
	\mathcal{P}
	=\ &
	\frac{e^2}{h}
	\bigg\{
	2D^2\kB T
\nonumber\\[-5pt]
	&
	+
	D(1-D)
	\!
	\sum\limits_{n=-\infty}^{+\infty}
	\!
	\big[n\hbar\Omega+e(V_\LL-V_\RR)\big]
	\big|\Cc_{n}\big|^2
\nonumber\\[-2pt]
	&
	\hspace*{58pt}
	\times
	\coth\bigg(
	\!
	\frac{n\hbar\Omega+e(V_\LL-V_\RR)}{2\kB T}
	\!
	\bigg)
	\!
	\bigg\}
	,
\end{align}
for a system with a gate applied to a single edge channel and a constant voltage bias~\mbox{$V_\LL-V_\RR$}, see Sec.~\ref{sec:Model_single_channel}.

\section{\label{app:Deriv_C}Derivation of the $\Cc_n$ coefficients}

In this appendix, we derive explicit functions for the coefficents $\Cc_n$, occurring in the Floquet scattering matrix, Eq.~(\ref{eq:SFnc_A_def}). Using the expansion of the internal potential~$\dUc(t)$ into harmonic components, 
\begin{equation}
	\dUc(t)
	=
	\dVg
	\sum\limits_{k=1}^{\Kmax}
	\text{Re}
	\Big\{
	2\Fg_k
	\,
	\text{e}^{-ik\Omega t}
	\,
	\alphac(k\Omega)
	\Big\}
	,
\end{equation}
with the lever arm function~$\alphac(\omega)$ given by Eq.~(\ref{eq:alphac1}), one finds the phase~$\phi^{(\CC)}(t)$, Eq.~(\ref{eq:phieta}), to have the form
\begin{align}
\!\!
	\phi^{(\CC)}(t)
	=&
	-2\frac{|e|\dVg}{\hbar\Omega}
	\sum_{k=1}^{\Kmax}
	\frac{\sin(k\Omega\tauc/2)}{k}
\nonumber\\
	&\times
	\Big[
	\text{Re}\big\{2\Fg_k\alphac(k\Omega)\big\}
	\cos\!\big(k\Omega [t-\tauc/2]\big)
\nonumber\\
	&
	\hspace*{5pt}
	+
	\text{Im}\big\{2\Fg_k\alphac(k\Omega)\big\}
	\sin\!\big(k\Omega [t-\tauc/2]\big)
	\Big]
	.
\end{align}
Then, after inserting the above expression into Eq.~(\ref{eq:Ceta}), and employing the Jacobi-Anger expansion~\cite{Abramowitz_book} for the factor~$\exp\!\big[\!-i\phi^{(\text{c})}(t)\big]$, namely, 
\begin{gather}
	\text{e}^{iz\cos\theta}
	=
	\!
	\sum\limits_{l=-\infty}^{+\infty}
	\!
	i^l
	J_l(z)
	\text{e}^{il\theta}
	,
\\
	\text{e}^{iz\sin\theta}
	=
	\!
	\sum\limits_{l=-\infty}^{+\infty}
	\!
	J_l(z)
	\text{e}^{il\theta}
	,
\end{gather}
where $J_l(z)$ stands for  the $l$th Bessel function of the first kind, one derives
\begin{align}\label{eq:coeffCc}
\!
	\Cc_n
	=&
	\int\limits_{0}^{\perT}
	\!
	\frac{\intd t}{\perT}
	\,
	\text{e}^{in\Omega t}
\nonumber\\
	&
	\times
	\prod\limits_{\substack{k=1\\[1.5pt] \text{Re}\{\!z_k\!\}\neq0}}^{\Kmax}
	\!\!
	\bigg\{
	\sum\limits_{l=-\infty}^{+\infty}
	\!
	i^l
	J_l\big(\text{Re}\{\!z_k\!\}\big)
	\text{e}^{il\Omega (t-\tauc/2)}
	\bigg\}
\nonumber\\
	&
	\times
	\prod\limits_{\substack{k^\prime=1\\[1.5pt] \text{Im}\{\!z_{k^\prime}\!\}\neq0}}^{\Kmax}
	\!\!
	\bigg\{
	\sum\limits_{l^\prime=-\infty}^{+\infty}
	\!\!
	J_{l^\prime}\big(\text{Im}\{\!z_{k^\prime}\!\}\big)
	\text{e}^{il^\prime\Omega (t-\tauc/2)}
	\bigg\}
	,
\end{align}
with
\begin{equation}
	z_k
	=
	2\frac{|e|\dVg}{\hbar\Omega}
	\cdot
	\frac{\sin(k\Omega\tauc/2)}{k}
	\cdot
	2\Fg_k\alphac(k\Omega)
	.
\end{equation}
\new{%
Equation~(\ref{eq:coeffCc}) is essential for calculating the number of excess emitted particles~$\dNeh$, Eqs.~(\ref{eq:Npart_1})-(\ref{eq:dNeh}).
}



%

\end{document}